\documentclass[a4paper,11pt]{article}
\pdfoutput=1 

\usepackage{jcappub} 
\usepackage{subcaption}
\usepackage[T1]{fontenc} 
\title{\boldmath On angular dependent response to gravitational-wave signals for time-delay interferometry combinations}
\author[a]{Pan-Pan Wang,}
\author[a]{Hao-Kang Chen,}
\author[b,c]{Wei-Liang Qian,}
\author[a]{Rui Luo,}
\author[a]{Jing Zhou,}
\author[a]{Wei-Sheng Huang,}
\author[a]{Yu-Jie Tan,}
\author[a]{and Cheng-Gang Shao}

\affiliation[a]{National Gravitation Laboratory, MOE Key Laboratory of Fundamental Physical Quantities Measurement, and School of Physics, Huazhong University of Science and Technology, Wuhan 430074, People's Republic of China}
\affiliation[b]{Escola de Engenharia de Lorena, Universidade de S\~ao Paulo, 12602-810, Lorena, SP, Brazil}
\affiliation[c]{Faculdade de Engenharia de Guaratinguet\'a, Universidade Estadual Paulista, 12516-410, Guaratinguet\'a, SP, Brazil}

\emailAdd{ppwang@hust.edu.cn}
\emailAdd{d202280025@hust.edu.cn}
\emailAdd{wlqian@usp.br}
\emailAdd{ruiluo@hust.edu.cn}
\emailAdd{18782530063@163.com}
\emailAdd{huangws22@hust.edu.cn}
\emailAdd{yjtan@hust.edu.cn}
\emailAdd{cgshao@hust.edu.cn}

\abstract{Space-based gravitational wave (GW) detectors are designed for wave sources in the millihertz band with different locations and orientations.
	Time-delay interferometry (TDI) technique is an indispensable ingredient in space-borne GW detection that effectively suppresses the laser phase noise.
	The abundant TDI solutions derived in the literature also feature distinct angular-dependent sensitivities.
	Because a GW source's angular location is unknown prior to the signals' detection, a solid-angle average is often performed when analyzing the sensitivity function of a given TDI combination.
	The present study explores the angular dependence of the detector's sensitivity.
	This detail is relevant, because once the initial detection is achieved, the source's location can be extracted and used to provide information on a refined TDI combination tailored for the specific GW source.
	As the TDI technique is a post-processing algorithm, such a procedure can be implemented in practice.
	We evaluate the angular dependence of the detector's response function to the GW signals for different TDI combinations as a function of the orientation angles.
	Moreover, we classify the response functions into seven categories at the low-frequency limit, leveraging the characteristics of the underlying geometrical TDI combinations.
	By further averaging out the azimuthal angle $\phi_D$ in the detector's plane, the main features of the resulting response functions and their zenithal dependence with respect to the GW source are scrutinized.
	The findings presented in this work provide pertinent insights for ongoing space-borne detector programs.}
\begin{document}

\maketitle
\flushbottom


\section{Introduction}\label{section1} 

The detection of gravitational waves (GWs) by the LIGO and Virgo interferometers~\citep{LIGO-01, LIGO-02, LIGO-03, LIGO-04, LIGO-05, LIGO-06, LIGO-07, LIGO-08} has inaugurated a new era in astrophysics, enabling unprecedented insights into the Universe through this new means. 
These ground-based detectors, including LIGO~\citep{gw-ligo1,gw-ligo2}, Virgo~\citep{gw-virgo}, and KAGRA~\citep{gw-KAGRA1,gw-KAGRA2}, are sensitive to high-frequency GWs, predominantly from stellar-mass binary mergers, in the range of 10 Hz to $10^4$Hz. 
Despite their successes, a vast spectrum of sources, such as galactic binaries, supermassive black hole binaries, and stochastic early universe backgrounds, are associated with GWs in the low-frequency domain. 
The exploration of these signals necessitates the development of space-based detectors such as LISA~\citep{gw-lisa1,gw-lisa2}, TianQin~\citep{gw-tianqin}, and TaiJi~\citep{gw-Taiji}, which are designed to operate in the millihertz band.

The ongoing space-borne detectors, consisting of three spacecraft forming a nearly equilateral triangle, measure GW-induced Doppler shifts in the laser light exchanged between the spacecraft. 
However, unlike their ground-based counterparts, these detectors face unique challenges due to their armlength variations and the influence of solar system gravity. 
In this regard, time-delay interferometry (TDI), first proposed by Tinto and Armstrong, plays a critical role in mitigating these challenges, particularly laser phase noise~\citep{tdi-01,tdi-02,tdi-03}. 
Over the past two decades, various algebraic~\citep{tdi-geome-2002,tdi-laser-04, Wu:2022bys, Qian:2022twp, Wu:2023key} and geometric approaches~\citep{tdi-geometric-2005, tdi-geometric-2020, tdi-geometric-2022} for obtaining the first and second-generation TDI solutions have been proposed and developed. 
The algebraic methods aim to derive the TDI combinations by solving the algebraic TDI equations, while geometric approaches are essentially exhaustive search algorithms whose solutions possess intuitive interpretations regarding the laser beams' virtual space-time trajectories.
These advancements have significantly enhanced the ability of space-based detectors to suppress laser phase noise and achieve the required sensitivity for the target GWs. 
The second-generation TDI combinations, particularly, meet the stringent performance criteria necessary for these ambitious empirical endeavors.

While the TDI algorithm primarily eliminates laser phase noise, each specific TDI combination inevitably imparts its unique footprint on the processed signals and residual noise due to its algebraic nature. 
Consequently, the abundant TDI solutions in the literature lead to diverse angular-dependent response functions and sensitivity curves.
A GW source's angular location might be unknown prior to the signals' detection, and therefore, a solid-angle average is often performed when analyzing the sensitivity function of a given TDI combination~\citep{response-full-03,response-full-04}.
The present study is motivated to study the angular dependence of the detector's sensitivity.
Such a perspective is relevant because once the initial detection is achieved, the source's location can be extracted and used to furnish information on a refined optimal combination tailored for the specific GW source.
As the TDI technique is a post-processing algorithm, such a procedure can be implemented in practice.
We evaluate the angular dependence of the detector's response function, as a function of the solid angles, to the GW signals for different TDI combinations.
At the low-frequency limit, we classify the response functions into seven categories, leveraging the characteristics of the underlying geometrical TDI combinations.
The properties, particularly the angular distribution of the obtained sensitivity curves of all forty-five geometric TDI combinations up to sixteen links, are analyzed.
To analyze the signal-to-noise ratio (SNR) for different TDI combinations, the signal needs to be integrated over the source trajectory.
For the LISA detector, the orbital period is one year, and for high-SNR sources that can be identified in a short time, a short data segment can be selected, during which $\theta_D$ and $\phi_D$ remain nearly unchanged in the detector's frame. 
For the TianQin detector, the angle $\theta_T$ is fixed, while $\phi_T$ has a period of 3.65 days.
Averaging over $\phi_T$ is equivalent to integrating over the source's trajectory, enabling an analysis that leads to an analytical expression for the response function, which depends on $\theta_T$.
Since space-based GW detection mainly focuses on the millihertz frequency band, this provides angular dependence diagrams for different TDI combinations in the low-frequency approximation.
These diagrams allow for the analysis of the response strength of the same TDI combination for different $\theta_D$ and $\phi_D$.
For the TianQin detector, the analysis further integrates over $\phi_T$, enabling the study of the response dependence on different $\theta_T$ values. 
From the analytical expression, it is straightforward to determine the values of $\theta_T$ that result in maximum and minimum responses.

The remainder of the paper is organized as follows.
In Sec.~\ref{section2}, we review the main features of the space-based GW detector and the implementation of the TDI algorithm.
In Sec.~\ref{section3}, we discuss the coordinate transform between the barycentric and the detector frames.
In Sec.~\ref{section4}, we elaborate on the general formalism of the angular distributions of the response functions.
The characteristics of the obtained response functions for specific TDI combinations of interest are analyzed in Sec.~\ref{section5}.
The angular dependence of the response functions of the LISA and TianQin detectors in the barycentric frame is analyzed in Sec.~\ref{section6}.
The last section is devoted to concluding remarks.
In Appendix ~\ref{threeTerms}, we present a summary of three relevant combinations of TDI coefficients for all forty-five geometric TDI solutions.

\section{The space-based GW detector and TDI algorithm}\label{section2}

In this section, we briefly review the layout of the space-borne GW detector and the relevant interferometric data streams.
The implementation of the TDI algorithm for suppressing the laser phase noise is also elaborated.

\subsection{The space-based GW detector and data streams}\label{section2.1}

\begin{figure}[tbp]
	\centering
	\includegraphics[width=\linewidth]{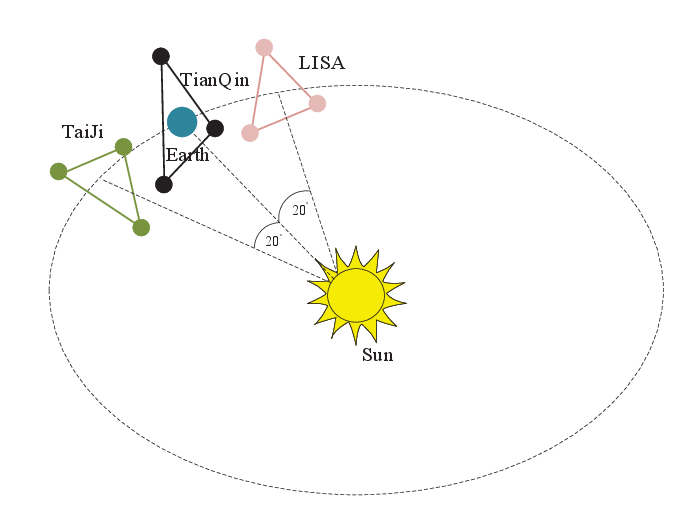}
	\caption{\label{fig1} The spatial layout of the three ongoing space-based GW detectors.}
\end{figure}

A typical space-based GW detector consists of three spacecraft forming a nearly equilateral triangle, as depicted in Fig.~\ref{fig1}.
In particular, for LISA (TaiJi) detector, the spacecraft are situated approximately $20^{\circ}$ behind (ahead) of Earth in the heliospheric orbit, with an inclination angle of $60^{\circ}$ between the plane of the triangular constellation and the ecliptic plane. 
The three spacecraft maintain the geometry of an equilateral triangle with side lengths of $L=2.5\times {{10}^{9}}\rm{m}$ and $3\times {{10}^{9}}\rm{m}$, respectively.
For TianQin, the spacecraft rotates in a geocentric orbit around Earth and revolves around the Sun with an armlength of $L=1.7\times {{10}^{8}}\rm{m}$.
The detector plane's normal vector points to the binary system RX J0806.3+1527.
Therefore, the main differences between space-based GW detectors' constellations reside in the orientation of the detector plane, the armlength, and the orbital period of the spacecraft.

\begin{figure}[tbp]
	\centering
	\includegraphics[width=\linewidth]{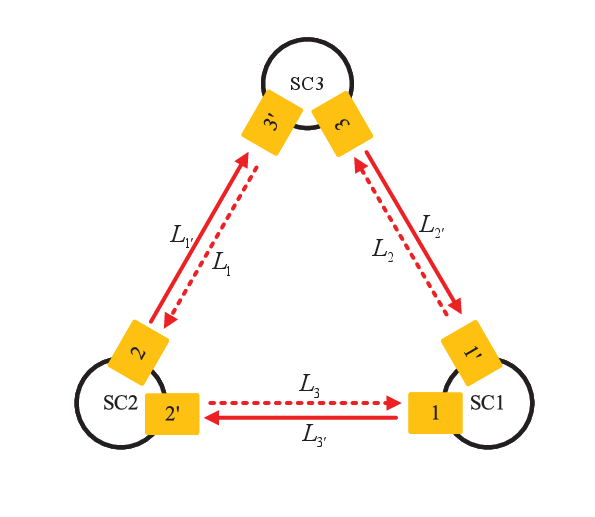}
	\caption{\label{fig2} An illustration of a space-based detector's various arm lengths~\citep{tdi-03}.}
\end{figure}

For the convenience of analysis, we establish the following conventions in this paper.
The three spacecraft of the detector are labeled as SC$i$ $(i=1, 2, 3)$, where each spacecraft carries two nearly identical optical benches marked by $i$ and ${i}'$.
The armlength opposite satellite $i$ is referred to as ${L}_{i}$ (${L}_{i'}$), regarding the laser beam propagates in the counterclockwise (clockwise) direction, as illustrated in Fig.~\ref{fig2}.
It is straightforward to eliminate the entire optical bench noise and laser phase noise on the primed optical benches by properly recombining the raw data streams.
The resulting data streams read~\citep{tdi-03}

\begin{subequations}
	\begin{align}
	& {\eta }_{i}(t)=h_i(t)+{D}_{i-1}p_{i+1}(t)-{{p}_{i}}(t)+\frac{2\pi}{\lambda_{(i+1)'}}{{{\vec{n}}}_{i-1}}\left[ {{D}_{i-1}}{{{\vec{\delta }}}_{(i+1{)}'}}(t)-{{{\vec{\delta }}}_{i}}(t) \right]+N_{i}^{\rm{opt}}(t), \label{etai}\\ 
	& {\eta }_{i'}(t)=h_{i'}(t)+{D}_{(i+1)'}{{p}_{i-1}}(t)-{{p}_{i}}(t)+\frac{2\pi}{\lambda_{(i-1)}}{{{\vec{n}}}_{i+1}}\cdot \left[ \vec \delta _{i'}(t)- {D}_{(i+1)'}{\vec \delta_{i-1}}(t) \right]+N_{i'}^{\rm{opt}}(t),\label{etaip} 
	\end{align}
\end{subequations}
where ${\eta }_{i} $ is known as the science data, it represents the laser beam emitted from SC$(i+1)$ interferes with the local laser on SC$i$ and therefore associated with a delay along the armlength $L_{i-1}$.
Similarly, the data stream ${\eta }_{i'} $ corresponds to the laser emitted from SC$(i-1)$ experiences a delay related to $L_{(i+1)'}$ interferes with the local laser on SC$i$ to generate a beat.
These interferometric measurements are affected by the GW signal ${h}_{i}$ along various arms, the laser phase noise ${p}_{i}$ associated with the lasers installed on the three optical benches, test mass vibrational noise 
$\delta _{i}$, the wavelength of the laser $\lambda_i$, and the optical path noise $ N_{i}^{\rm opt}$.
In the above expressions, ${D}_{i}$ denotes the time-delay operator, which is defined as follows when applied to an arbitrary data stream: 
 \begin{align}\label{delayd}
 	D_i x(t) =& x\left(t - \frac{L_i(t)}{c}\right),\\\notag
 	{ D_j}{ D_i}x(t) =& { D_j}x\left(t - \frac{L_i(t)}{c}\right) = x\left(t - \frac{L_i\left(t - \frac{L_j(t)}{c}\right)}{c} - \frac{L_j(t)}{c}\right).
 \end{align}
Successive application of multiple time-delay operators can be denoted as follows:
\begin{align}\label{delaycon}
D_{i}D_{j}x\left( t \right)=D_{ij}x\left( t \right).
\end{align}
For the various types of noise present in Eqs.~\eqref{etai}-\eqref{etaip}, the test mass noise and optical path noise are considered inevitable, which we regard as the noise floor.
The TDI technique suppresses the laser phase noise, which is much more significant than the GW signal.
In this paper, we will not consider clock noise and tilt-to-length noise, among other factors, by simply stating that they can be mitigated using pertinent techniques~\citep{tdi-clock4,ttl-2016}.

\subsection{The TDI technique}\label{section2.1}

To suppress the laser phase noise, the data streams defined in Eqs.~\eqref{etai}-\eqref{etaip} can be further combined as:
\begin{align}\label{TDIequ}
		\text{TDI}=\sum\limits_{i=1,2,3}\left(P_i\eta_{i}+P_{i'}\eta_{i'}\right),
\end{align}
where the coefficients ${{P}_{i}},{P}_{i'}$ are polynomials in the time delay operator.

\begin{figure}[tbp]
	\centering
	\includegraphics[width=\linewidth]{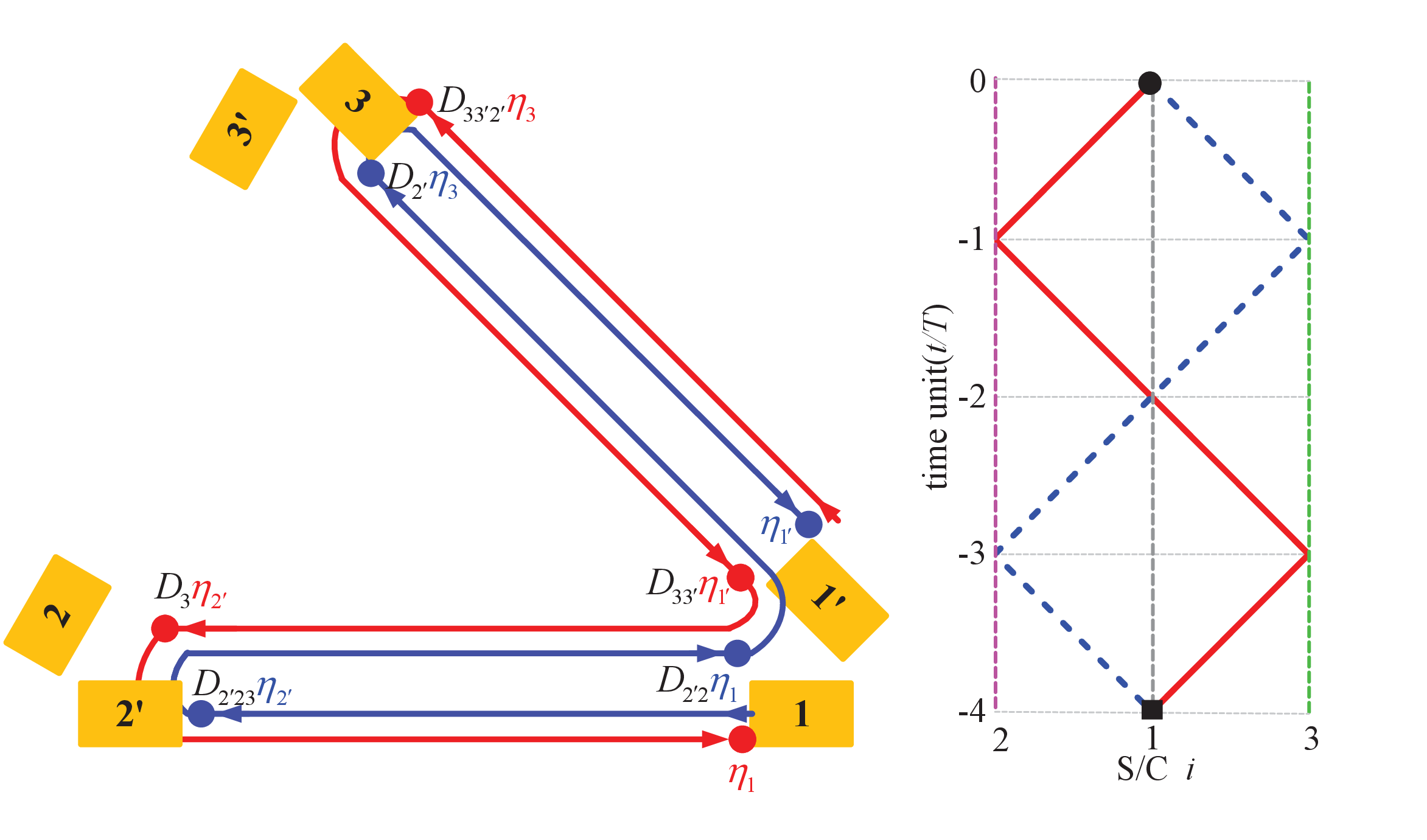}
	\caption{\label{fig3} 
		An illustration of the Michelson TDI combination from a geometric TDI perspective~\citep{tdi-geometric-2022}. 
		Left: The virtual laser beams' trajectories that furnish the Michelson combination. 
		Right: The equivalent space-time diagram of the combination. 
		While the colors of the trajectories remain unchanged, one uses dashed lines to represent propagations in the negative direction of time.
	}
\end{figure}

As an example, the first-generation Michelson TDI combination ${X}_{1}$ is given by~\citep{tdi-03}
\begin{equation}
	\begin{aligned}
		{{X}_{1}}=\left( -1+{ D_{{2'}\text{2}}} \right){{\eta }_{1}}+\left( { D}_{{2'}}- D_{33'2'} \right){{\eta }_{3}}+\left( \text{1}-D_{33'} \right){{\eta }_{{1'}}}+\left( D_{2'23}-D_3 \right){\eta }_{2'} .\label{TDIX1}
	\end{aligned}
\end{equation}
By comparing Eq.~\eqref{TDIX1} against Eq.~\eqref{TDIequ}, the polynomial coefficients are found to be
\begin{equation}
	\begin{aligned}
	& P_1=-1+ D_{2'2}, \\ 
	& P_2=0, \\ 
	& P_3= D_{2'}- D_{33'2'}, \\ 
	& P_{1'}=1-D_{33'}, \\ 
	& P_{2'}={D}_{2'23}- D_3, \\ 
	& P_{3'}=0. \\ \label{PnX1}
	\end{aligned}
\end{equation}

As a geometric TDI solution, the laser beams' virtual trajectories of the Michelson combination $X_1$ are presented in Fig.~\ref{fig3}, where one assumes that only the two arms between SC1-SC2 and SC1-SC3 are functional.
Through a specific TDI combination, such as Eq.~\eqref{PnX1}, the laser phase noise cancels out in the expression Eq.~\eqref{TDIequ}.
However, as mentioned before, the resulting GW signals and residual noise are also recombined in terms of the polynomials of the delay operators $P_i$.
Subsequently, the abundant TDI solutions established in the literature give rise to diversified response functions with different angular dependencies.

\section{The relationship between the barycentric and detector frames} \label{section3}

To facilitate the analysis of the angular response function in different coordinate systems, we will introduce the coordinate transform between the barycentric and detector frames.
For a source located in a given direction $\theta_B,\phi_B$ in the barycentric frame, its corresponding trajectory in the detector frame is represented by $\left( {\theta _D}(t),{\phi _D}(t) \right)$.
The relationships between the barycentric and detector frames, for the LISA and TianQin detectors, are detailed below.

\subsection{The coordinate transform between the barycentric and the LISA detector frames}\label{section3.1}

LISA lags the Earth by $20^{\circ}$ along its orbit around the Sun, with the plane of the LISA detector inclined at an angle of $60^{\circ}$ to the ecliptic plane, and its orbital period is 1 year.
The solar barycentric coordinate system is described using a Cartesian coordinate system $(x_B, y_B, z_B)$, and the LISA detector coordinate system is denoted as $(x_L, y_L, z_L)$ .
The transformation from the barycentric coordinate system to the LISA detector coordinate system is as follows: the vector ${\vec r_B} =(x_B,y_B,z_B)$  is transformed to ${\vec r_L} =(x_L,y_L,z_L)$ through the matrix $R$, i.e.~\citep{optimal-SNR-2003},
\begin{align}\label{LISArotation}
\left( {\begin{array}{*{20}{c}}
{{x_L}}\\
{{y_L}}\\
{{z_L}}
\end{array}} \right) = {R_z}\left( {{\psi _c}} \right){R_x}\left( {\frac{\pi }{3}} \right){R_z}\left( {{\psi _a}} \right)\left( {\begin{array}{*{20}{c}}
{{x_B}}\\
{{y_B}}\\
{{z_B}}
\end{array}} \right),
\end{align}
where
\begin{align}
\begin{array}{l}
{R_z}\left( {{\psi _a}} \right) = \left( {\begin{array}{*{20}{c}}
{\cos {\psi _a}}&{\sin {\psi _a}}&0\\
{ - \sin {\psi _a}}&{\cos {\psi _a}}&0\\
0&0&1
\end{array}} \right),\\
{R_x}\left( {\frac{\pi }{3}} \right) = \left( {\begin{array}{*{20}{c}}
1&0&0\\
0&{\frac{1}{2}}&{\frac{{\sqrt 3 }}{2}}\\
0&{ - \frac{{\sqrt 3 }}{2}}&{\frac{1}{2}}
\end{array}} \right),\\
{R_z}\left( {{\psi _c}} \right) = \left( {\begin{array}{*{20}{c}}
{\cos {\psi _c}}&{\sin {\psi _c}}&0\\
{ - \sin {\psi _c}}&{\cos {\psi _c}}&0\\
0&0&1
\end{array}} \right),
\end{array}
\end{align}
Here, ${\psi _a} = \omega t + {\alpha _0}$, ${\psi _c} =  - \omega t + {\beta _0}$, and $\omega  = \frac{{2\pi }}{{{T_ \odot }}}$. $T_ \odot$ is the orbital period of the LISA detector,  ${\alpha _0}$ and ${\beta _0}$ are constants fixing initial conditions when the observation begins, which is set to 0 in this paper.
Based on Eqs.~\eqref{LISArotation}, the relationship between the solar barycentric system and the LISA detector coordinate system can be given as~\citep{optimal-SNR-2003}
\begin{align}\label{LISAbary}
\sin{\theta _L}\cos {\phi _L} =& \frac{{\sqrt 3 }}{2}\cos {\theta _B}\sin {\psi _c} + \sin {\theta _B}\sin {\phi _B}\left( {\cos {\psi _c}\sin {\psi _a} + \frac{1}{2}\cos {\psi _a}\sin {\psi _c}} \right)\\\notag
 +& \sin {\theta _B}\cos {\phi _B}\left( {\cos {\psi _a}\cos {\psi _c} - \frac{1}{2}\sin {\psi _a}\sin {\psi _c}} \right),\\\notag
\sin {\theta _L}\sin {\phi _L} =& \frac{{\sqrt 3 }}{2}\cos {\theta _B}\cos {\psi _c} + \cos {\phi _B}\sin {\theta _B}\left( { - \cos {\psi _a}\sin {\psi _c} - \frac{1}{2}\cos {\psi _c}\sin {\psi _a}} \right)\\\notag
 +& \sin {\theta _B}\sin {\phi _B}\left( { - \sin {\psi _a}\sin {\psi _c} + \frac{1}{2}\cos {\psi _a}\cos {\psi _c}} \right),\\\notag
\cos {\theta _L} =& \frac{1}{2}\cos {\theta _B} - \frac{{\sqrt 3 }}{2}\sin {\theta _B}\sin \left( {{\phi _B} - {\psi _a}} \right).
\end{align}

In the barycentric coordinate system, for a fixed wave source direction $\left( {{\theta _B},{\phi _B}} \right)$, the apparent source direction in the detector coordinate system changes with time.
As time passes, $\left( {{\theta _L},{\phi _L}} \right)$ varies, and the source is moving in the LISA reference frame.

\begin{figure}[tbp]
	\centering
	\includegraphics[width=\linewidth]{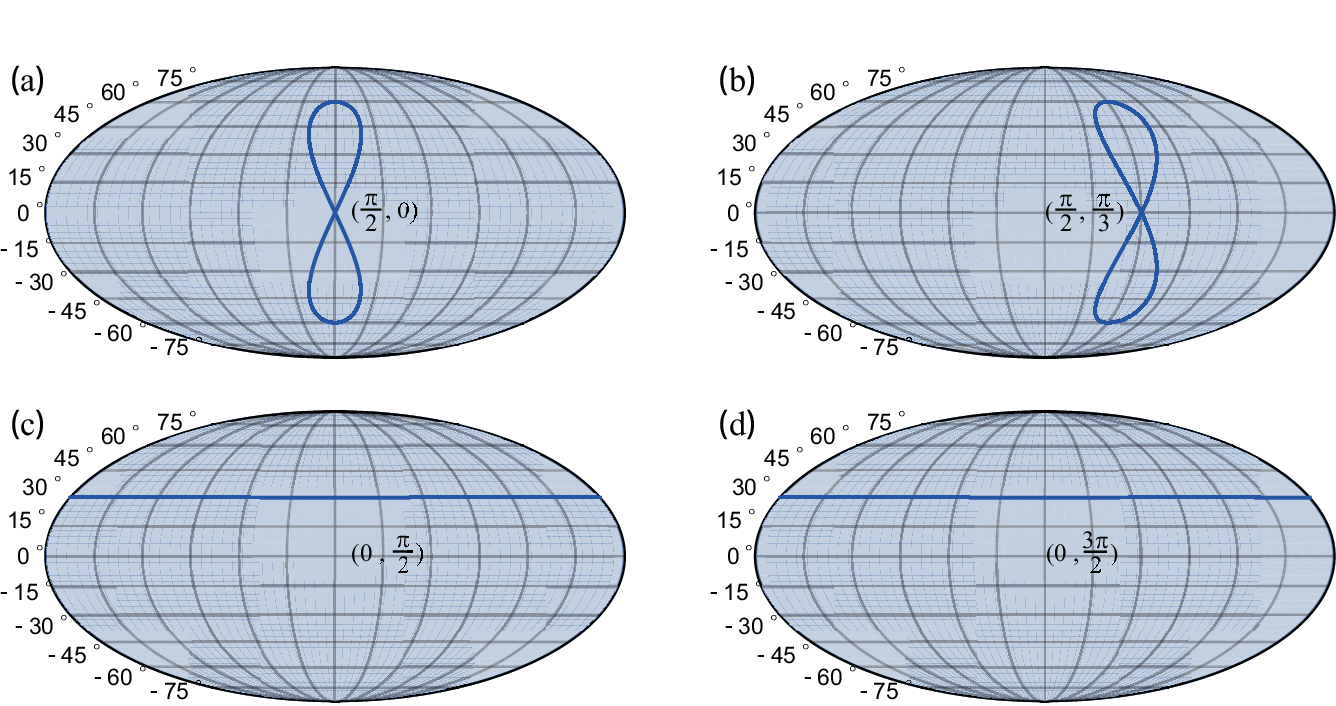}
	\caption{\label{thetafix} 
		The source trajectory in the LISA reference frame when $(\theta_B,\phi_B)$ are set to 
		$\left( {\frac{\pi }{2},0} \right),\left( {\frac{\pi }{2},\frac{\pi }{3}} \right),
		\left( {0,\frac{\pi }{2}} \right),\left( {0,\frac{{3\pi }}{2}} \right)$.
	}
\end{figure}

Based on Eqs.~\eqref{LISAbary}, we can describe the trajectory of $(\theta_L,\phi_L)$ in the two-dimensional plane for a fixed source.
When the inclination angle $\theta_B$ of the wave source is fixed and $\phi_B$ varies, the shape of the source trajectory remains the same in the LISA detector.
For example, Fig.~\ref{thetafix}  illustrates the source trajectory in the LISA detector when $(\theta_B,\phi_B)$ are set to $\left( {\frac{\pi }{2},0} \right),\left( {\frac{\pi }{2},\frac{\pi }{3}} \right),\left( {0,\frac{\pi }{2}} \right)$, and $\left( {0,\frac{{3\pi }}{2}} \right)$.
When $\theta_B=\frac{\pi }{2}$, the source trajectory in the LISA detector coordinate system forms a figure-eight shape, while for $\theta_B= 0$, the trajectory becomes a straight line.

\begin{figure}[tbp]
	\centering
	\includegraphics[width=\linewidth]{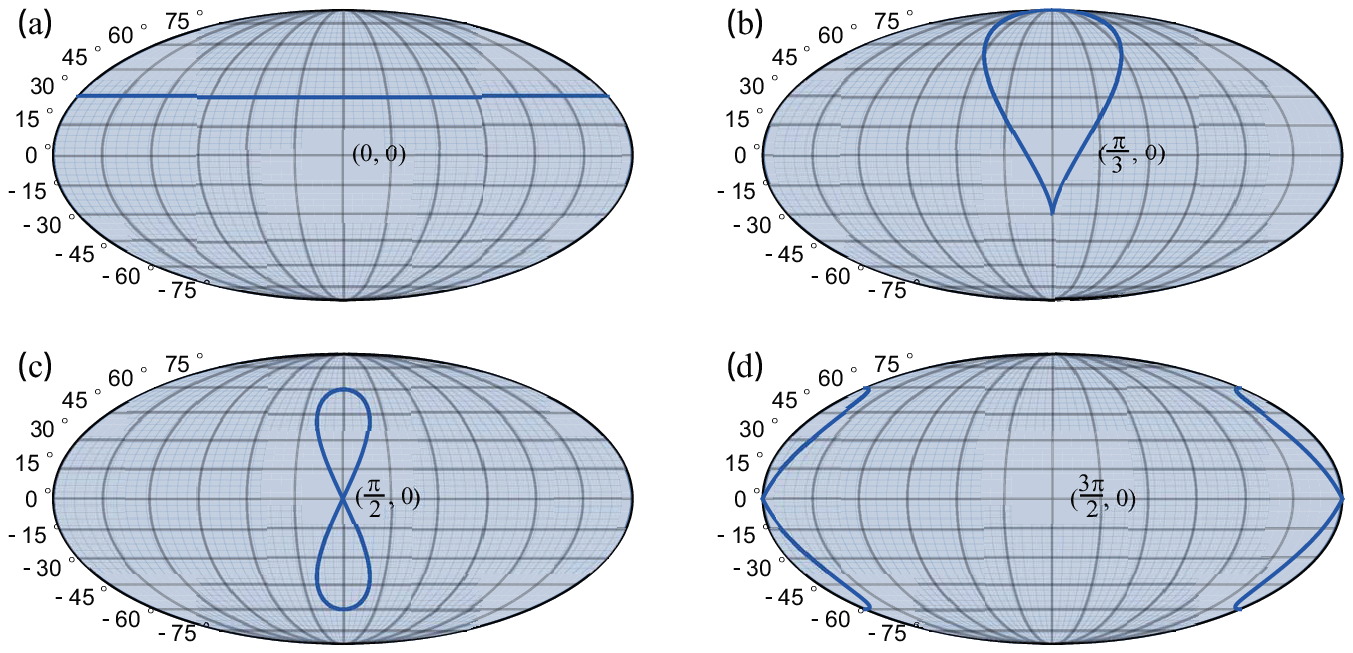}
	\caption{\label{phifix} 
		The source trajectory in the LISA reference frame when 
		$\left( {0,0} \right),\left( {\frac{\pi }{3},0} \right),
		\left( {\frac{\pi }{2},0} \right),\left( {\frac{{3\pi }}{2},0} \right)$.
	}
\end{figure}

When the inclination angle $\phi_B$ of the wave source is fixed and $\theta_B$ varies, the source trajectory in the LISA detector changes.
As shown in Fig.~\ref{phifix} , the source trajectory $(\theta_B,\phi_B)$ in the LISA detector is illustrated for different values of $\left( {0,0} \right),\left( {\frac{\pi }{3},0} \right),\left( {\frac{\pi }{2},0} \right)$, and $\left( {\frac{{3\pi }}{2},0} \right)$.

\subsection{The coordinate transform between the barycentric and the TianQin detector frames}\label{section3.2}

\begin{figure}[tbp]
	\centering
	\includegraphics[width=\linewidth]{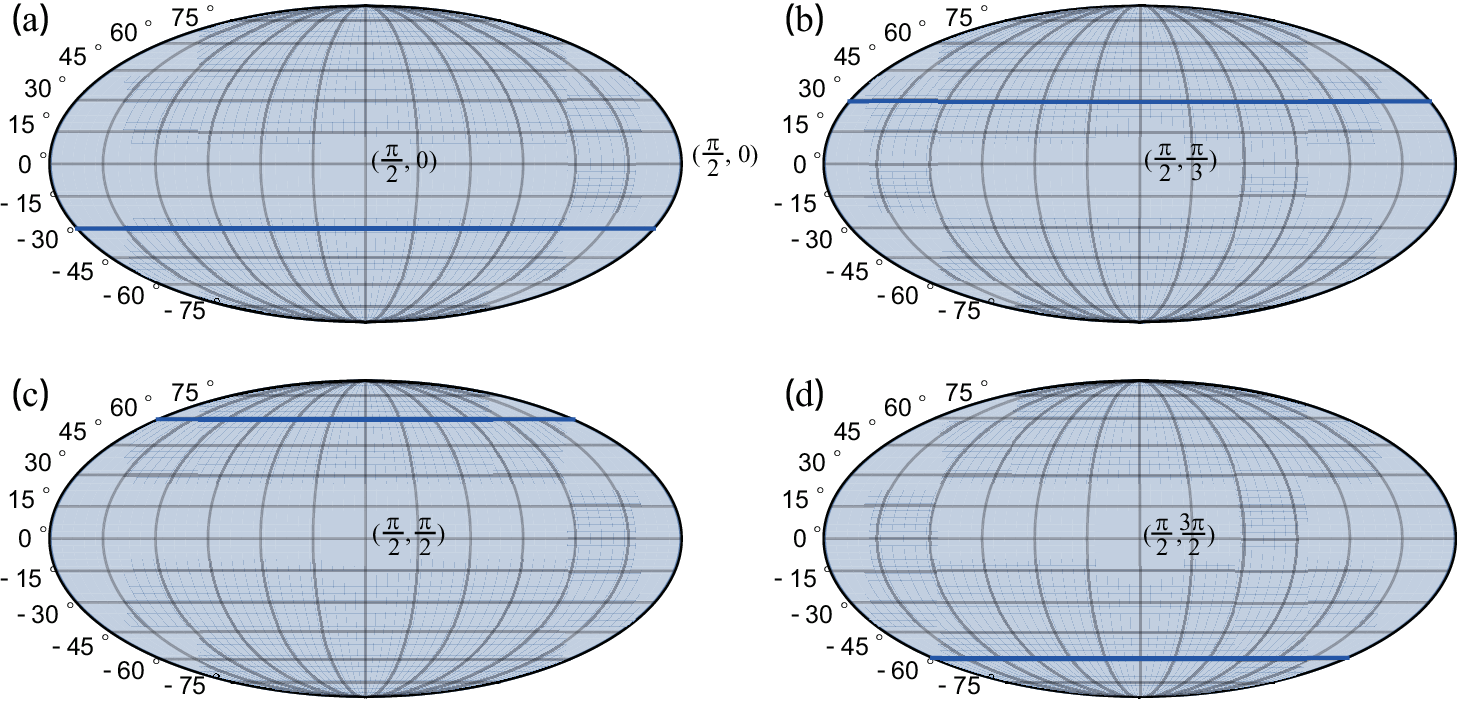}
	\caption{\label{TQ-thetaphi} 
		The source trajectory in the TianQin detector when 
		$\left( {\frac{\pi }{2},0} \right),\left( {\frac{\pi }{2},\frac{\pi }{3}} \right),
		\left( {0,\frac{\pi }{2}} \right),\left( {0,\frac{{3\pi }}{2}} \right)$.
	}
\end{figure}

Throughout the entire mission, the TianQin detector's plane always points towards its reference source, J0806.3+1527. 
The reference source J0806.3+1527 has a latitude and longitude $\left( {\theta _J},{\phi _J} \right)$ in the heliocentric ecliptic coordinate system, specifically ${\theta _J} = {94.7^{\circ} },{\phi _J} = {120.5^{\circ} }$ .
The transformation from the barycentric coordinate system to the TianQin detector coordinate system is as follows: the vector ${\vec r_B} =(x_B,y_B,z_B)$ is transformed to ${\vec r_T} =(x_T,y_T,z_T)$  through the matrix $R$, i.e.,
\begin{align}
\left( {\begin{array}{*{20}{c}}
{{x_T}}\\
{{y_T}}\\
{{z_T}}
\end{array}} \right) = {R_z}\left( {{\varphi _c}} \right){R_y}\left( {{\theta _J}} \right){R_z}\left( {{\phi _J}} \right)\left( {\begin{array}{*{20}{c}}
{{x_B}}\\
{{y_B}}\\
{{z_B}}
\end{array}} \right),
\end{align}
where
\begin{align}\label{TQrotation}
\begin{array}{l}
{R_z}\left( {{\phi _J}} \right) = \left( {\begin{array}{*{20}{c}}
{\cos {\phi _J}}&{\sin {\phi _J}}&0\\
{ - \sin {\phi _J}}&{\cos {\phi _J}}&0\\
0&0&1
\end{array}} \right),\\
{R_y}\left( {{\theta _J}} \right) = \left( {\begin{array}{*{20}{c}}
{\cos {\theta _J}}&{\rm{0}}&{ - \sin {\theta _J}}\\
{\rm{0}}&{\rm{1}}&0\\
{\sin {\theta _J}}&0&{\cos {\theta _J}}
\end{array}} \right),\\
{R_z}\left( {{\varphi _c}} \right) = \left( {\begin{array}{*{20}{c}}
{\cos {\varphi _c}}&{\sin {\varphi _c}}&0\\
{ - \sin {\varphi _c}}&{\cos {\varphi _c}}&0\\
0&0&1
\end{array}} \right).
\end{array}
\end{align}
At time $t=0$, one finds
\begin{align}
\sin {\theta _T}\cos {\phi _{_T}}\left( 0 \right) =& \cos {\theta _J}\cos \left( {{\phi _B} - {\phi _J}} \right)\sin \theta {,_B} - \cos {\theta _B}\sin {\theta _J},\\\notag
\sin {\theta _T}\sin {\phi _T}\left( 0 \right) =& \sin {\theta _B}\sin \left( {{\phi _B} - {\phi _J}} \right),\\\notag
\cos {\theta _T}\left( 0 \right) =& \cos {\theta _B}\cos {\theta _J} + \cos \left( {{\phi _B} - {\phi _J}} \right)\sin {\theta _B}\sin {\theta _J},
\end{align}
After a certain period of time, we have
\begin{align}\label{TQbary}
\sin {\theta _T}\cos {\phi _T} =& \sin {\theta _T}\cos \left[ {{\phi _{_T}}\left( 0 \right) + {\varphi _c}} \right]\\\notag
 =& \cos {\theta _J}\cos \left( {{\phi _B} - {\phi _J}} \right)\sin {\theta _B}\cos {\varphi _c} - \cos {\theta _B}\sin {\theta _J}\cos {\varphi _c} + \sin {\theta _B}\sin \left( {{\phi _B} - {\phi _J}} \right)\sin {\varphi _c},\\\notag
\sin {\theta _T}\sin {\phi _T} =& \sin {\theta _T}\sin \left[ {{\phi _{_T}}\left( 0 \right) + {\varphi _c}} \right]\\\notag
 =&  - \cos {\theta _J}\cos \left( {{\phi _B} - {\phi _J}} \right)\sin {\theta _B}\sin {\varphi _c} + \cos {\theta _B}\sin {\theta _J}\sin {\varphi _c} + \sin {\theta _B}\sin \left( {{\phi _B} - {\phi _J}} \right)\cos {\varphi _c},\\\notag
\cos {\theta _T} =& \cos {\theta _T}\left( 0 \right) = \cos {\theta _B}\cos {\theta _J} + \cos \left( {{\phi _B} - {\phi _J}} \right)\sin {\theta _B}\sin {\theta _J},
\end{align}
Here, ${\varphi _c} = \omega t$.
Based on Eqs.~\eqref{TQbary}, one can describe the trajectory of $(\theta_T,\phi_T)$ in the two-dimensional plane for a fixed source.
In the solar barycentric coordinate system, for a fixed wave source direction $(\theta_B,\phi_B)$, the apparent source direction in the TianQin detector coordinate system changes over time.
Both $\theta_T,\phi_T$ vary with time, but since the normal direction remains fixed, its trajectory forms concentric circles around J0806.3+1527.
Regardless of the value of the wave source direction $\theta_B,\phi_B$, the trajectory in the TianQin detector always appears as a straight line in the two-dimensional plane.
Fig.~\ref{TQ-thetaphi} illustrates the source trajectory in the TianQin detector for different values of $\left( {\frac{\pi }{2},0} \right),\left( {\frac{\pi }{2},\frac{\pi }{3}} \right),\left( {0,\frac{\pi }{2}} \right),\left( {0,\frac{{3\pi }}{2}} \right)$.
Therefore, for the TianQin detector, the integration over the trajectory is essentially an integration over $\phi_T$.

\section{Angular dependent response functions to GW signals for an arbitrary TDI combination}\label{section4}

Recently, we analyzed all forty-five geometric TDI combinations up to sixteen links by employing a ternary exhaustive search algorithm and analyzed the obtained sensitivity curves~\citep{Sensitivity-2023}.
It is found that there are a total of eleven categories regarding distinct sensitivity curves. 
However, such an analysis was carried out by averaging over the orientation, namely, the angles $(\theta, \phi)$, to the GW source.
Nonetheless, even though a GW source’s angular location might be unknown prior to the signals’ detection, the information becomes available once the initial detection is achieved.
Such information can be utilized to provide a refined optimal combination tailored for the specific GW source. 
As the TDI technique is a post-processing algorithm, such a procedure can be implemented in practice to improve sensitivity. 
Based on the above considerations, we explore the angular distribution of the detector’s response function to the GW signals for different TDI combinations as a function of the orientation angles. 

\subsection{The angular dependence of the response functions}\label{section4.1}

\begin{figure}[tbp]
	\centering
	\includegraphics[width=\linewidth]{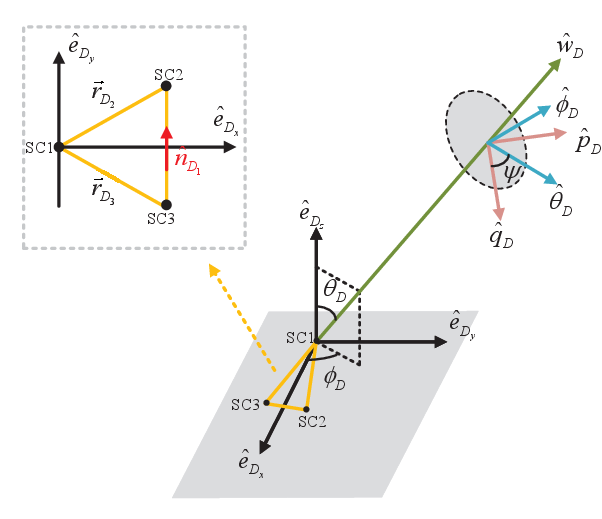}
	\caption{\label{detecreferframe} 
		The chosen coordinate system in the detector's frame utilized for the calculations~\citep{response-full-03}.
	}
\end{figure}

We will use the detector frame to evaluate the Doppler shifts induced by the GW signals.
In such a frame of reference, the adopted coordinate system is illustrated in Fig.~\ref{detecreferframe}.
The origin of the coordinate system is placed at the location of SC1.
The $x-y$ plane is chosen to coincide with that of the three spacecraft, where ${{\hat{e}}_{x_D}}$ equally divides the angle formed by SC1-SC2 and SC1-SC3.
The azimuthal and zenithal angles $(\theta_D, \phi_D)$ measure the orientation of the GW source, $\hat{w_D}$, as indicated in Fig.~\ref{detecreferframe}.
Subsequently, the position and unit vectors of the three detectors are given by 
\begin{equation}\label{vecn}
		\begin{aligned}
	& {{{\vec{r}}}_{D_1}}=(0,0,0),{{{\hat{n}}}_{D_1}}=(0,1,0), \\ 
	& {{{\vec{r}}}_{D_2}}=-L{{{\hat{n}}}_{D_3}}=L\left(\cos \frac{\gamma }{2},\sin \frac{\gamma }{2},0\right), \\ 
	& {{{\vec r}}_{D_3}}=L{{{\hat{n}}}_{D_2}}=L\left(\cos \frac{\gamma }{2},-\sin \frac{\gamma }{2},0\right), \\ 
	\end{aligned}
\end{equation}
where the subscript ``$D$'' denotes the LISA or TianQin frame, represented by ``$L$'' or ``$T$'' respectively.

Two additional coordinate systems~\citep{gwdirection-01} are also relevant: the observational reference frame (ORF) in terms of $\left(\hat{\theta }_D,\hat{\phi }_D,\hat{w}_D\right)$ and the canonical reference frame (CRF) denoted by $\left(\hat{p}_D,\hat{q}_D,\hat{w}_D\right)$.
The ORF measures the orientation of the source in terms of the unit vectors of the azimuthal and zenithal angles, $\hat{\theta}_D,\hat{\phi}_D$, and the radial direction $\hat{w}_D$.
It is straightforward to find the relations
\begin{equation}\label{hatw}
\begin{aligned}
\hat{w}_D=-\hat{k}_D=\sin \theta_D \cos \phi_D {{\hat{e}}_{D_x}}+\sin \theta_D \sin \phi_D{{\hat{e}}_{D_y}}+\cos \theta_D {{\hat{e}}_{D_z}},
\end{aligned}
\end{equation}
\begin{equation}\label{theta}
\hat{\theta }_D=\partial \hat{w}_D/\partial \theta_D =\cos \theta_D \cos \phi_D {{\hat{e}}_{x}}+\cos \theta_D \sin \phi_D {{\hat{e}}_{D_y}}-\sin \theta_D {{\hat{e}}_{D_z}},
\end{equation}
and
\begin{equation}\label{phi}
\hat{\phi }_D=\partial \hat{w}_D/\left(\sin \theta_D \partial \phi_D \right)=-\sin \phi_D {{\hat{e}}_{D_x}}+\cos \phi_D{{\hat{e}}_{D_y}},
\end{equation}
where $\hat{k}_D$ is the direction of the wave vector.
On the other hand, the CRF is more suitable for describing the polarization state from the source's perspective.
It can be attained by simply rotating the observational reference frame clockwise around the ${\hat{w}}$-axis by an angle $\psi$, forming two new orthogonal vectors 
\begin{equation}
	\begin{aligned}
	& \hat{p}_D=\cos \psi \hat{\theta }_D-\sin \psi \hat{\phi }_D, \\ 
	& \hat{q}_D=\sin \psi \hat{\theta }_D+\cos \psi\hat{\phi }_D, \\ 
	\end{aligned}
\end{equation}
where the orthogonal basis $\hat{p}_D$ and $\hat{q}_D$ can be readily used to write down the polarization tensors for the plus and cross modes in a fashion that is irrelevant to the detector's motion
\begin{equation}
\begin{aligned}
	e_{ij}^{+}=\hat{p}_{D_i}\hat{p}_{D_j}-\hat{q}_{D_i} \hat{q}_{D_j},\\ 
	e_{ij}^{\times }=\hat{p}_{D_i}\hat{q}_{D_j}+\hat{q}_{D_i}\hat{p}_{D_j}.\\ 
\end{aligned}
\end{equation}

To evaluate the response function, let us first consider a simple scenario of one-way transmission, where a laser beam emitted at SC$i$ propagates in the $\hat{n}_D$ direction along arm $L$ and is then received at SC$j$.
The presence of the GW signal produces an additional relative Doppler shift, which is expressed as follows~\citep{tdi-laser-04}
\begin{equation}
	{{\tilde{s}}_{L}}\equiv \frac{\delta v(\Omega )}{{{v}_{0}}}=\frac{\tilde{h}(\Omega )}{2(1-\hat{k}_D\cdot \hat{n}_D)}{{e}^{i\Omega \frac{L+\hat{k}_D\cdot {\vec  r}_{D_{i}}}{c}}}\left[ 1-{{e}^{-i\frac{\Omega L}{c}(1-\hat{k}_D\cdot \hat{n}_D)}} \right],\label{oneArmSig}
\end{equation}	
where  ${\nu }_0$ is the center frequency of the laser,
$\delta \nu(\Omega )$ gives the frequency shift owing to the presence of GW,
$\Omega =2\pi f$ is the circular frequency of the GW,
$L$ is the armlengths connecting spacecraft SC$i$ and SC$j$ located at $\vec{r}_{D_i}$ and $\vec{r}_{D_j}$.
The amplitude of the GW is $\tilde{h}(\Omega )$ is expressed in the frequency domain, which contains the sum of all possible polarization states, namely
\begin{equation}
	\tilde{h}(\Omega )=\sum_A \tilde{h}_A(\Omega ){{\xi }_{\hat{n};A}},
\end{equation}
where the polarization state $A={+,\times}$, and $\xi_{\hat{n}_D;A}$ is the direction function defined as
\begin{equation}\label{denfi}
\begin{aligned}
	&{\xi }_{\hat{n}_D;+}={{(\hat{\theta }_D\cdot {{\hat{n}_D}})}^{2}}-{{(\hat{\phi }_D\cdot {{\hat{n}_D}})}^{2}},\\ 
	&{\xi }_{\hat{n}_D;\times }=2(\hat{\theta }_D\cdot {{\hat{n}_D}})(\hat{\phi }_D\cdot {{\hat{n}_D}}).\\ 
\end{aligned}
\end{equation}

One can generalize Eq.~\eqref{oneArmSig} to a given TDI combination involving a specific detector layout.
Also, for $\tilde{h}_A(\Omega)$, it is convenient to separate the part corresponding to specific GW strain from that concerning the spatial layout of the detector and the specific TDI combinations.
For the data streams acquired by the optical benches $i$ and $i'$, we have
\begin{equation}
	\begin{aligned}
	& {\tilde{s}}_{i}=F_{\eta_i,A}(\Omega)\tilde{h}_A(\Omega), \\ 
	& {\tilde{s}}_{i'}=F_{\eta_{i'},A}(\Omega)\tilde{h}_A(\Omega), \\ \label{etaRes}
	\end{aligned}
\end{equation}
where the antenna response functions $F_{\eta_i,A}(\Omega)$ and $F_{\eta_{i'},A}(\Omega)$ for the polarization state $A=(+,\times)$ are given by
\begin{equation}
	\begin{aligned}
	& F_{{\eta }_{i,A}}(\Omega )=\frac{e^{i\Omega (L_{i-1}-\hat{w}_D\cdot {\vec r}_{D_{i+1}})/c}}{2(1+\hat{w}_D\cdot \hat{n}_{D_{i-1}})}\left[ 1-e^{-i\Omega L_{i-1}(1+\hat{w}_D\cdot {\hat n}_{D_{i-1}})/c} \right]{\xi _{i-1; A}}, \\ 
	& {{F}_{{{\eta }_{{{i}'}}},A}}(\Omega )=\frac{{{e}^{i\Omega ({{L}_{i+1}}-\hat{w}_D\cdot {{{\vec{r}}}_{D_{i-1}}})/c}}}{2(1-\hat{w}_D\cdot {{{\hat{n}}}_{D_{i+1}}})}\left[ 1-{{e}^{-i\Omega {{L}_{i+1}}(1-\hat{w}_D\cdot {{{\hat{n}}}_{D_{i+1}}})/c}} \right]{{\xi }_{i+1; A}}, \\ 
	\end{aligned}
\end{equation}
where one introduces a shorthand for the direction function ${\xi }_{i;A}\equiv {\xi }_{\hat{n}_{D_i};A}$ and $\hat{n}_i$ gives the direction of $L_i$.

Substituting Eqs.~\eqref{vecn},\eqref{theta},\eqref{phi} into Eq.~\eqref{denfi}, 
in the detector frame, one may write down the explicit forms of the direction functions as follows
\begin{equation}\label{xiplus}
\begin{aligned}
{\xi _{1; + }} &= {\cos \theta_D}^2 {\sin \phi_D}^2  - {\cos \phi_D}^2 ,\\
{\xi _{2; + }}& = {\cos \theta_D}^2 {\cos \tilde \phi_D}^2  - {\sin\tilde \phi_D}^2,\\
{\xi _{3; + }} &= {\cos \theta_D}^2 {\cos \tilde \phi_D}^2 -{\sin\underset{\tilde{}}{\phi_D}}^2,\\
\end{aligned}
\end{equation}
and
\begin{equation}\label{xicross}
\begin{aligned}
{\xi _{1; \times }} &= \cos \theta_D \sin2\phi_D ,\\
{\xi _{2; \times }} &=  - \cos \theta_D \sin2\tilde \phi_D ,\\
{\xi _{3; \times }} &=  - \cos \theta_D \sin2\underset{\tilde{}}{\phi_D}.\\
\end{aligned}
\end{equation}
Also, we have
\begin{equation}\label{wn}
\begin{aligned}
\hat w_D \cdot {{\hat n}_{D_1}} &= \sin \theta_D \sin \phi_D,\\
\hat w_D \cdot {{\hat n}_{D_2}} &= \sin \theta_D \cos \tilde \phi_D,\\
\hat w_D \cdot {{\hat n}_{D_3}}& =  - \sin \theta_D \cos\underset{\tilde{}}{\phi_D},\\ 
\end{aligned}
\end{equation}
where $\underset{\tilde{}}{\phi_D}= \phi_D  - \frac{\gamma }{2},\tilde \phi_D=\phi_D  + \frac{\gamma }{2}$, and in the remainder of this paper we take $\gamma=\frac{\pi}{3}$ as a reasonable approximation.

By substituting Eq.~\eqref{etaRes} into the definition of a TDI solution Eq.~\eqref{TDIequ}, one finds that the response function of the polarization state $A$ for a given TDI combination is given by
\begin{equation}\label{respofu}
	{{F}_{A}}=\sum\limits_{i=1}^{3}{{{{\tilde{P}}}_{i}}{{F}_{{{\eta }_{i}};}}_{A}}+{{\tilde{P}}_{{{i}'}}}{{F}_{{{\eta }_{{{i}'}}};}}_{A} ,
\end{equation}
where the angular dependence enters through the relative orientation between the detector and GW source governed by Eqs.~\eqref{xiplus},~\eqref{xicross}, and~\eqref{wn}.

\subsection{Angular dependent response function for an arbitrary TDI combination}\label{section4.2}

To evaluate the response function of the resulting GW signals, one proceeds to calculate the square module of Eq.~\eqref{respofu}, which gives
\begin{equation}\label{FA}
	4{{\left| {{F}_{A}} \right|}^{2}}={\left|a_1{\xi }_{1;A}+a_2{\xi }_{2;A}+ a_3{\xi }_{3;A} \right|}^{2},
\end{equation}
where
\begin{equation}
\begin{aligned}
	&a_1={{\tilde{P}}_{2}}{{e}^{-iu\sin \theta_D \cos \tilde{\phi }_D}}\frac{1-{{e}^{-iu(1+\sin \theta_D \sin \phi_D )}}}{1+\sin \theta_D \sin \phi_D }+{{\tilde{P}}_{{{3}'}}}{{e}^{-iu\sin \theta_D \cos \underset{\tilde{}}{{\phi}_D}}}\frac{1-{{e}^{-iu(1-\sin \theta_D \sin \phi_D )}}}{1-\sin \theta_D \sin \phi_D },\\
	&a_2={{\tilde{P}}_{3}}\frac{1-{{e}^{-iu(1+\sin \theta_D \cos \tilde{\phi_D })}}}{1+\sin \theta_D \cos \tilde{\phi_D }}+{{\tilde{P}}_{{{1}'}}}{{e}^{-iu\sin \theta_D \cos \tilde{\phi_D }}}\frac{1-{{e}^{-iu(1-\sin \theta_D \cos \tilde{\phi_D })}}}{1-\sin \theta_D \cos \tilde{\phi_D }},\\
 &a_3={{\tilde{P}}_{1}}{{e}^{-iu\sin \theta_D \cos \underset{\tilde{}}{\phi_D}}}\frac{1-{{e}^{-iu(1-\sin \theta_D \cos \underset{\tilde{}}{\phi_D})}}}{1-\sin \theta_D \cos \underset{\tilde{}}{\phi_D}}+{{\tilde{P}}_{{{2}'}}}\frac{1-{{e}^{-iu(1+\sin \theta_D \cos \underset{\tilde{}}{\phi_D})}}}{1+\sin \theta_D \cos \underset{\tilde{}}{\phi_D}},
\end{aligned}
\end{equation}
where $u=\frac{2\pi f L}{c}$ is a dimensionless quantity, 
${{\tilde{P}}_{i}}$ denotes the Fourier transform of the coefficient $P_i$, 
where the Fourier transform of the time-delay operator is ${\tilde{ \cal D}} = e^{iu}$.

In the context of space-borne GW detection, the focus is on the millihertz frequency band. 
Therefore, apart from the last subsection, we will primarily adopt the low-frequency approximation $u\ll 1$ as a reasonable estimation.
At the low-frequency limit, for most TDI combinations encountered in the literature~\citep{tdi-geometric-2005, tdi-geometric-2020, tdi-geometric-2022}, one has $\tilde{P}_i=O(u)$ and $\tilde{P}_i+\tilde{P}_{(i+1)'}=O(u^2)$.
Subsequently, the resulting expressions for $a,b,c$ can be written as
\begin{equation}\label{appabcMod}
\begin{aligned}
	&a_1=iu\left[ {{{\tilde{P}}}_{2}}+{{{\tilde{P}}}_{{{3}'}}}-\frac{iu}{2}\left({{{\tilde{P}}}_{2}}-{{{\tilde{P}}}_{{{3}'}}}\right)\sin \theta_D \left(\sin \phi_D +\cos \tilde{\phi }_D-\cos \underset{\tilde{}}{\phi_D}\right) \right]+O(u^5)
    =iu\left( {{{\tilde{P}}}_{2}}+{{{\tilde{P}}}_{{{3}'}}} \right)+O(u^5),\\
	&a_2=iu\left( {{{\tilde{P}}}_{3}}+{{{\tilde{P}}}_{{{1}'}}} \right)\left[1-\frac{iu}{2}\left(1+\sin \theta_D \cos \tilde{\phi }_D\right)\right]+O(u^5),\\
 &a_3= iu\left( {{{\tilde{P}}}_{1}}+{{{\tilde{P}}}_{{{2}'}}} \right)\left[1-\frac{iu}{2}\left(1+\sin \theta_D \cos \underset{\tilde{}}{\phi_D}\right)\right] +O(u^5).
\end{aligned}
\end{equation}
By substituting Eq.~\eqref{appabcMod} into Eq.~\eqref{FA}, the response function gives
\begin{equation}\label{responappabc}
\begin{aligned}
&\frac{4}{{{u}^{2}}}{{\left| F_{+} \right|}^{2}}_{\text{TDI}}\underset{u \ll 1}{\mathop{\to }}\,{{\left| \begin{aligned}
			& ({{{\tilde{P}}}_{1}}+{{{\tilde{P}}}_{{{2}'}}})\left( {{\cos }^{2}}\theta_D {{\cos }^{2}}\underset{\tilde{}}{\phi_D}-\sin^{2}\underset{\tilde{}}{\phi_D}\right)+({{{\tilde{P}}}_{2}}+{{{\tilde{P}}}_{{{3}'}}})\left( {{\cos }^{2}}\theta_D {{\sin }^{2}}\phi_D -{{\cos }^{2}}\phi_D  \right) \\ 
			& +({{{\tilde{P}}}_{3}}+{{{\tilde{P}}}_{{{1}'}}})\left( \cos^{2}\theta_D {{\cos }^{2}}\tilde{\phi }_D-\sin^2\tilde{\phi }_D \right) \\ 
		\end{aligned} \right|}^{2}},\\
&\frac{4}{{{u}^{2}}}{{\left| F_{\times } \right|}^{2}}_{\text{TDI}}\underset{u \ll 1}{\mathop{\to }}\,{{\cos }^{2}}\theta_D {{\left| -({{{\tilde{P}}}_{1}}+{{{\tilde{P}}}_{{{2}'}}})\sin2\underset{\tilde{}}{\phi_D}+({{{\tilde{P}}}_{2}}+{{{\tilde{P}}}_{{{3}'}}})\sin2\phi_D -({{{\tilde{P}}}_{3}}+{{{\tilde{P}}}_{{{1}'}}})\sin2\tilde{\phi }_D \right|}^{2}}.
\end{aligned}
\end{equation}

One observes that the results given by Eq.~\eqref{responappabc} are valid for most scenarios.
The resulting response function is in order $O(u^4)$.
However, for the particular category of fully-symmetric Sagnac combinations, their polynomial coefficients satisfy the limit ${{\tilde{P}}_{i}}+{{\tilde{P}}_{(i+1{)}'}}= o(u^3)$.
In this regard, the expansions in Eqs.~\eqref{appabcMod} do not suffice. 
One has to proceed further to the third order, which gives 
\begin{equation}\label{fullyappabcMod}
\begin{aligned}
    &a_1 = iu\left[ \tilde{P}_{2}\frac{(iu)^2}{6}\sin \theta_D \sin \phi_D + \left( \tilde{P}_{2}+\tilde{P}_{3'} \right) \right] + O(u^5),\\
    &a_2 = iu\left[ \tilde{P}_{3}\frac{(iu)^2}{6}\sin \theta_D \cos \tilde{\phi}_D + \left( \tilde{P}_{3}+\tilde{P}_{1'} \right) \right] + O(u^5), \\
    &a_3 = iu\left[-\tilde{P}_{1}\frac{(iu)^2}{6}\sin \theta_D \cos \underset{\tilde{}}{\phi_D} + \left( \tilde{P}_{1}+\tilde{P}_{2'} \right) \right] + O(u^5).
\end{aligned}
\end{equation}
Subsequently, the corresponding response functions of the fully-symmetric Sagnac combinations for the plus and cross modes are of the order $O(u^5)$, which read
\begin{equation}\label{responfullyappabc}
\begin{aligned}
	&\frac{4}{{{u}^{2}}}{{\left| F_{+} \right|}^{2}}_{\text{TDI-}\zeta }\underset{u \ll 1}{\mathop{\to }}\,{{\left( \frac{{{(iu)}^{2}}}{6}\sin \theta_D  \right)}^{2}}{{\left| \begin{aligned}
			& -{{{\tilde{P}}}_{1}}\cos \underset{\tilde{}}{\phi_D}\left( {{\cos }^{2}}\theta_D {{\cos }^{2}}\underset{\tilde{}}{\phi_D}-\sin^2\underset{\tilde{}}{\phi_D } \right)+{{{\tilde{P}}}_{2}}\sin \phi  \\ 
			& \left( {{\cos }^{2}}\theta_D {{\sin }^{2}}\phi_D -{{\cos }^{2}}\phi_D  \right) +{{{\tilde{P}}}_{3}}\cos \tilde{\phi }_D\left( \cos^{2}\theta_D {{\cos }^{2}}\tilde{\phi }_D-\sin^2\tilde{\phi }_D \right) \\ 
		\end{aligned} \right|}^{2}},\\
&\frac{4}{{{u^2}}}{\left| {F_ \times} \right|^2}_{{\rm{TDI - }}\zeta }\mathop  \to \limits_{u \ll 1} {\cos ^2}\theta_D {\left( {\frac{{{{(iu)}^2}}}{6}\sin \theta_D } \right)^2}{\left| { - {{\tilde P}_1}\cos\underset{\tilde{}}{\phi_D} \sin2\underset{\tilde{}}{\phi_D} + {{\tilde P}_2}\sin \phi_D \sin2\phi_D  + {{\tilde P}_3}\cos \tilde \phi_D \sin2\tilde \phi_D } \right|^2}.
\end{aligned}
\end{equation}

The analytic results given by Eqs.~\eqref{responappabc} and~\eqref{responfullyappabc} essentially cover all the forty-five geometric TDI combinations up to sixteen links.
The response to GW signals depends on the azimuthal and zenithal angles $(\theta,\phi)$ and the specific TDI combination through the coefficients $\tilde{P}_i$.
The specific features of the resulting response functions will be further explored below in Sec.~\ref{section5}.

\subsection{The noise PSD}\label{section4.3}

To accurately estimate the detector's capability, it is crucial to understand both its response function and the associated noise characteristics.
In the subsection, we explore the residual noise the power spectral density (PSD).
Upon substituting Eqs.~\eqref{etai} and \eqref{etaip} into Eq.~\eqref{TDIequ}, the TDI technique effectively suppresses the laser phase noise below the noise floor, resulting in the following residual test mass noise terms:
\begin{equation}\label{tmnoise}
    \text{TDI}^{a}=\sum_{i=1}^{3}\frac{2\pi}{ \lambda}\left\{ \begin{aligned}
        & -\left[ P_{i}+P_{(i+1)'}\mathcal{D}_{(i-1)'} \right]\vec n_{(i-1)}\vec \delta_{(i)}(t) +\left[P_{i-1}\mathcal{D}_{(i+1)}+P_{i'} \right]\vec n_{(i+1)}\vec \delta_{(i)'}(t) \\ 
    \end{aligned} \right\}.
\end{equation}
The higher-order effects resulting from the wavelength $\lambda$ were ignored in this context.
The corresponding noise PSD is
\begin{equation}\label{tmnoisepsd}
    S_{\text{TDI}^a}=S_a(\Omega) \sum_{i=1}^{3}\left[ \left| \tilde{P}_{i}(\Omega) + \tilde{P}_{(i+1)'}(\Omega)\tilde{\mathcal{D}}_{i-1}(\Omega) \right|^2 + \left| \tilde{P}_{i'}(\Omega) + \tilde{P}_{i-1}(\Omega)\tilde{\mathcal{D}}_{i+1}(\Omega) \right|^2 \right], 
\end{equation}
where $S_a=\frac{s_a^2 L^2}{u^2 c^4}$ represents the energy spectral density of the test mass noise. 
Here, $s_a$ denotes the amplitude spectral density (ASD) of the test-mass acceleration noise, predominantly present in the low-frequency range. 

The residual terms of the optical path noise are given by
\begin{equation}\label{opnoise}
    \text{TDI}^x = \sum_{i=1}^{3}\left(P_{i}N_{i}^{\text{opt}}(t) + P_{i'}N_{i'}^{\text{opt}}(t)\right).
\end{equation}
The corresponding noise PSD is
\begin{equation}\label{opnoisepsd}
    S_{\text{TDI}^x}(\Omega) = S_{\text{opt}}(\Omega) \sum_{i=1}^{3}\left[ |\tilde{P}_{i}(\Omega)|^2 + |\tilde{P}_{i'}(\Omega)|^2 \right],
\end{equation}
where $S_{\text{opt}}=\frac{u^2 s_x^2}{L^2}$ is the optical path noise energy spectral density, and $s_x$ is the ASD of aggregate optical-path noise, also referred to as displacement noise or readout noise, primarily constituting the high-frequency component. 
According to Eqs.~\eqref{tmnoisepsd} and \eqref{opnoisepsd}, we observe that the noise PSD for various TDI combinations does not depend on the space orientation angle and, therefore, does not play a role in the angular dependence of the resulting sensitivity curves.

\section{Angular dependent response functions for the forty-five geometric TDI combinations}\label{section5}

In this section, we apply the results in the previous section to all the forty-five geometric TDI combinations up to sixteen links. 
Analyzing the properties of the obtained response functions shows that the response functions can be divided into seven categories at the low-frequency limit.
The properties of specific categories are analyzed.

\subsection{Specific TDI combinations at the low-frequency limit}\label{section5.1}

By observing Eqs.~\eqref{responappabc}, it is found that the response functions depend on the polynomial coefficients exclusively through three specific combinations $\tilde{P}_i+\tilde{P}_{(i+1)'}$ where $i=1, 2, 3$.
Therefore, one only needs to evaluate this three-tuple to assess the response function for a specific TDI combination at the low-frequency limit.
To comprehensively assess the TDI combination, we carry out such an analysis by enumerating all forty-five geometric TDI solutions up to sixteen links.

As an example, for the modified second-generation TDI combination $\left[ X \right]_{1}^{16} $, the Fourier transform of polynomial coefficients read
\begin{equation}\label{X1four}
	\begin{aligned}
	& {{\tilde{P}}}_{1}=1-{{e}^{2iu}}-{{e}^{4iu}}+{{e}^{6iu}}, \\ 
	& {{\tilde{P}}}_{2}=0, \\ 
	& {{\tilde{P}}}_{3}=-{{e}^{iu}}+{{e}^{3iu}}+{{e}^{5iu}}-{{e}^{7iu}}, \\ 
	& {{\tilde{P}}}_{{{1}'}}=-1+{{e}^{2iu}}+{{e}^{4iu}}-{{e}^{6iu}}, \\ 
	& {{\tilde{P}}}_{{{2}'}}={{e}^{iu}}-{{e}^{3iu}}-{{e}^{5iu}}+{{e}^{7iu}}, \\ 
	& {{\tilde{P}}}_{{{3}'}}=0, \\ 
\end{aligned}
\end{equation}
where one adopts the approximation of equal armlengths.
At low frequencies, one expands all the coefficients to the second order in $u$ and evaluates the three combinations
\begin{equation}
\begin{aligned}
	& n_a\equiv {{{\tilde{P}}}_{1}}+{{{\tilde{P}}}_{{{2}'}}}\sim 16{{\left( iu \right)}^{2}}, \\ 
	& n_b\equiv {{{\tilde{P}}}_{2}}+{{{\tilde{P}}}_{{{3}'}}}\sim 0, \\ 
	& n_c\equiv {{{\tilde{P}}}_{3}}+{{{\tilde{P}}}_{{{1}'}}}\sim -16{{\left( iu \right)}^{2}}. \label{ThreeTupleDef}\\ 
\end{aligned}
\end{equation}
As the response functions are entirely governed by a three-tuple $(n_a, n_b, n_c)$ defined in Eq.~\eqref{ThreeTupleDef}, it is convenient to use the properties of the latter to classify the angular dependence of the former.
For the combination in question, $\left[ X \right]_{1}^{16}$, we have $(n_a, n_b, n_c)=\left( 16,  0,  -16 \right){\left( iu \right)}^{2}$.
Employing this convention, the tuples for all the forty-five geometric TDI combinations are evaluated and summarized in Tabs.~\ref{12linkexpression}-\ref{16linkexpression17} in the appendix.
By observing Tabs.~\ref{12linkexpression}-\ref{16linkexpression17}, all these TDI combinations can be divided into seven categories, which are summarized in Tab.~\ref{class}.

\begin{table}
	\centering
	\caption{Seven different categories of the response functions at the low-frequency limit.
    The three-tuple of a combination pertaining to a given category is evaluated up to a constant.
    The results are regarding all forty-five geometric TDI solutions up to sixteen links.}
	\newcommand{\tabincell}[2]{\begin{tabular}{@{}#1@{}}#2\end{tabular}}
	\renewcommand\arraystretch{4}
	\begin{tabular}{|c|c|c|}
		\hline
		\hline
		Category & Three-tuple & TDI combinations\\
		\hline
		$1$ & $\left( 1,     0,    -1 \right)$&$\begin{aligned}
			& \left[ \alpha  \right]_{1}^{12},\left[ \alpha  \right]_{2}^{12},\left[ U \right]_{1}^{14},\left[ U \right]_{2}^{14},\left[ EP \right]_{1}^{14},\left[ X \right]_{1}^{\text{16}},\left[ X \right]_{2}^{\text{16}}, \\ 
			& \left[ E \right]_{1}^{\text{16}},\left[ E \right]_{2}^{\text{16}},\left[ PE \right]_{1}^{\text{16}},\left[ PE \right]_{2}^{\text{16}},\left[ PE \right]_{3}^{\text{16}},\left[ T \right]_{1}^{\text{16}},\left[ T \right]_{2}^{\text{16}}, \\ 
			& \left[ T \right]_{3}^{\text{16}},\left[ T \right]_{4}^{\text{16}},\left[ T \right]_{5}^{\text{16}},\left[ T \right]_{6}^{\text{16}},\left[ T \right]_{7}^{\text{16}},\left[ T \right]_{8}^{\text{16}} \\ 
		\end{aligned}$\\
		\hline
		$2$ & $\left(-1,     1,    0 \right)$&$\left[ T \right]_{11}^{\text{16}}$\\
		\hline
		$3$ & $\left( 0,     1,    -1\right)$&$\begin{aligned}
			& \left[ P \right]_{1}^{\text{16}},\left[ P \right]_{2}^{\text{16}},\left[ PE \right]_{4}^{\text{16}},\left[ PE \right]_{5}^{\text{16}},\left[ PE \right]_{6}^{\text{16}}, \\ 
			& \left[ PE \right]_{7}^{\text{16}},\left[ PE \right]_{8}^{\text{16}},\left[ PE \right]_{9}^{\text{16}},\left[ PE \right]_{10}^{\text{16}},\left[ T \right]_{16}^{\text{16}} \\ 
		\end{aligned}$\\
		\hline
		$4$ & $\left( 2,     -1,    -1\right)$&$\left[ U \right]_{1}^{\text{16}},\left[ U \right]_{\text{2}}^{\text{16}},\left[ U \right]_{3}^{\text{16}}$\\
		\hline
		$5$ & $\left(4,    -1,    -3\right)$&$\left[ U \right]_{5}^{\text{16}},\left[ U \right]_{6}^{\text{16}}$\\
		\hline
		$6$ & $\left( 8,    -3,   -5 \right)$&$\left[ U \right]_{\text{4}}^{\text{16}}$\\
		\hline
		$7$ & $\left(0, 0, 0\right)$&$\left[ \alpha  \right]_{3}^{12},\left[ EP \right]_{2}^{14},\left[ T \right]_{9}^{\text{16}},\left[ T \right]_{12}^{\text{16}},\left[ T \right]_{14}^{\text{16}},\left[ T \right]_{15}^{\text{16}},\left[ T \right]_{18}^{16},\left[ T \right]_{19}^{16}$\\
		\hline
		\hline
	\end{tabular}
	\label{class}
\end{table}

Each row of Tab.~\ref{class} corresponds to a category.
The first six rows give rise to non-vanishing response functions, where the coefficient combination is of the order ${\tilde{P}}_{i}+{\tilde{P}}_{(i+1{)}'}= O(u^2)$.
The angular dependence of the response functions is identical for the combinations pertaining to a given category.
The first three categories are equivalent to a permutation of the spacecraft.
It is noted that the last row of Tab.~\ref{class} corresponds to a fully-symmetric Sagnac combination, which consists of eight TDI combinations.
These particular geometric solutions have the following low-frequency limit ${\tilde{P}}_{i}+{\tilde{P}}_{(i+1{)}'}= o(u^3)$, and as a result, the response function vanishes up to the third order.
In this case, higher-order expansion must be performed to assess the response function, as elaborated in the text.

\begin{figure}[tbp]
	\centering
	\includegraphics[width=\linewidth]{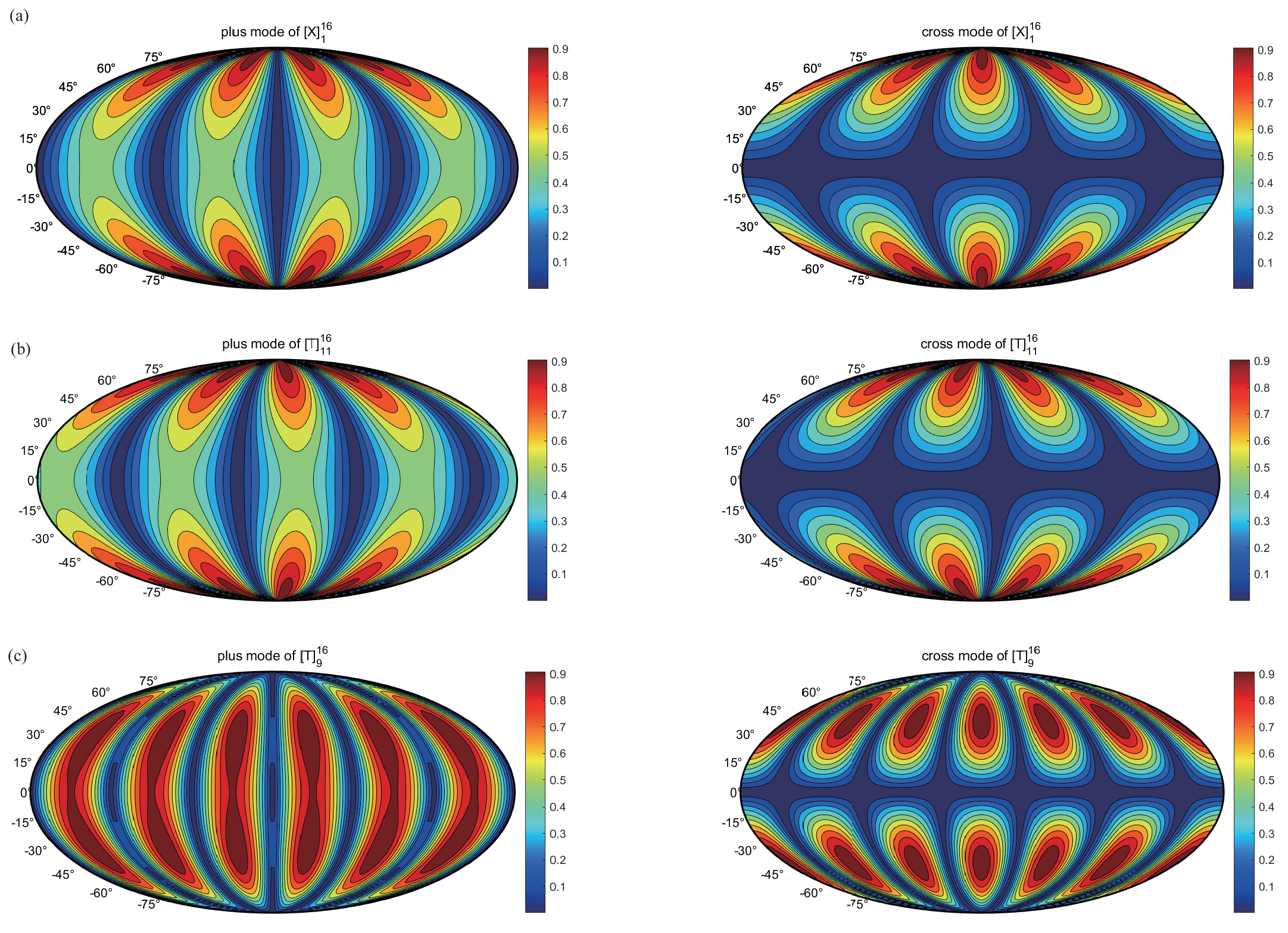}
	\caption{\label{LISAdetector} 
		(a) The angular dependence of the response function of the TDI combinations 
		$\left[ X \right]_{1}^{16}$ and $\left[ E \right]_{2}^{16}$ for the plus and cross modes. 
		(b) The angular dependence of the response function of the TDI combination 
		$\left[ T\right]_{11}^{16}$ for the plus and cross modes. 
		It is related to (a) by a rotation of $2\pi/3$ of the azimuthal angle $\phi$. 
		(c) The angular dependence of the response function of the TDI combination 
		$\left[ T \right]_{9}^{16}$ for the plus and cross modes.
	}
\end{figure}

\subsection{Angular dependent response functions}\label{section5.2}

Let us illustrate the results of the response function's angular distribution with a few numerical examples.
The first example concerns the modified second-generation Michelson combination $\left[ X \right]_{1}^{16}$.
By substituting the polynomial coefficient combinations $16\left(1, 0, -1\right){(iu)}^{2}$ from the second row of Tab.~\ref{16modlinkexpression} into Eq.~\eqref{responappabc}, one finds the corresponding angular dependence of the response functions for the plus and cross modes.
The obtained angular dependence belongs to category 1 listed on the second row of Tab.~\ref{class}, corresponding to the tuple $(1,0,-1)$.
The results are shown in Fig.~\ref{LISAdetector}(a).
Also, for the TDI combination $\left[ E \right]_{2}^{16}$, the coefficient combination $(2, 0, -2)(iu)^2$ indicates that it belongs to the same category associated with the tuple $(1,0,-1)$.
Therefore, the response function is essentially identical to Fig.~\ref{LISAdetector}(a).
For the TDI combination $\left[ T\right]_{11}^{16}$, it belongs to category 2 related to the tuple $(-1, 1, 0)$, and the resulting response function is given by Fig.~\ref{LISAdetector}(b).
A permutation between the spacecraft is equivalent to a rotation of $2\pi/3$ of the azimuthal angle $\phi$ on the detector's plane.
Lastly, for the fully-symmetric Sagnac TDI combination $\left[ T\right]_{9}^{16}$, the response function is presented in Fig.~\ref{LISAdetector}(c).
Even though the tuple vanishes, the strength of the response function is of a higher order, but the angular dependence persists.

\subsection{Angular dependent response functions for TianQin detector}\label{section5.3}
For the TianQin detector, on the other hand, $\phi_T$ evolves more rapidly.
For more extensive observations, we average the distribution in the azimuthal direction by integrating out $\phi_T$ and focus primarily on the dependence of the zenithal angle $\theta_T$. 

We use the components of the three-tuple, $({n}_{a},{n}_{b},{n}_{c})=\left( \tilde{P}_{1}+\tilde{P}_{2'},\tilde{P}_2+\tilde{P}_{3'},\tilde{P}_3+\tilde{P}_{1'} \right)$, to write down the response functions for a specific TDI solution.  
For the non-fully symmetric Sagnac combinations, integrating $\phi$ in Eq.~\eqref{responappabc} yields the responses for the plus and cross modes as follows 
\begin{equation}\label{responappabctheta}
	\begin{aligned}
		& \frac{4}{u^{2}}\left|F_{+}(\theta_T)\right|^{2}_{\text{TDI}}
		\underset{u\ll1}{\to}
		(n_{a}^{2}+n_{b}^{2}+n_{c}^{2})\left(\frac{3}{8}\cos^{4}\theta_T-\frac{1}{4}\cos^{2}\theta_T+\frac{3}{8}\right)
		\\[2pt]
		&\qquad\quad
		+2(n_{a}n_{b}+n_{b}n_{c}+n_{a}n_{c})\left(\frac{3}{16}\cos^{4}\theta_T-\frac{5}{8}\cos^{2}\theta_T+\frac{3}{16}\right) \\
		&\;=\left(\frac{3}{8}\cos^{4}\theta_T-\frac{1}{4}\cos^{2}\theta_T+\frac{3}{8}\right)
		\left[ n_{a}^{2}+n_{b}^{2}+n_{c}^{2}+n_{a}n_{b}+n_{b}n_{c}+n_{a}n_{c}\right]
		-\cos^{2}\theta_T\,(n_{a}n_{b}+n_{b}n_{c}+n_{a}n_{c}),  \\
		&\frac{4}{u^{2}}\left|F_{\times}(\theta_T)\right|^{2}_{\text{TDI}}
		\underset{u\ll1}{\to}
		\frac{1}{2}\cos^{2}\theta_T\left[ n_{a}^{2}+n_{b}^{2}+n_{c}^{2}-n_{a}n_{b}-n_{b}n_{c}-n_{a}n_{c}\right].
	\end{aligned}
\end{equation}

On the other hand, for the full-symmetric Sagnac combinations, one finds 
\begin{equation}
	\begin{aligned}
	& \frac{4}{{{u}^{2}}}{{\left| F_{+}(\theta_T ) \right|}^{2}}{{\left(\frac{6}{{{{\tilde{P}}}_{1}}{{(iu)}^{2}}}\right)}^{2}}\frac{1}{{{\sin }^{2}}\theta_T }\underset{u \ll 1}{\mathop{\to }}\,3\left(\frac{5}{16}{{\cos }^{4}}\theta_T -\frac{1}{8}{{\cos }^{2}}\theta_T +\frac{1}{16}\right) 
	+6\left(\frac{-7}{64}{{\cos }^{4}}\theta_T +\frac{5}{32}{{\cos }^{2}}\theta_T +\frac{1}{64}\right), \\ 
	&=\frac{9}{8}\left(\frac{\cos ^{2}\theta_T +1}{2}\right)^{2},\\
	 &\frac{4}{{{u}^{2}}}{{\left| F_{\times }(\theta_T ) \right|}^{2}}{{\left(\frac{6}{{{{\tilde{P}}}_{1}}{{(iu)}^{2}}}\right)}^{2}}\frac{1}{{{\sin }^{2}}\theta_T }\underset{u \ll 1}{\mathop{\to }}\frac{1}{4} \left(3+\frac{3}{2} \right){{\cos \theta_T }^{2}}=\frac{9}{8}{{\cos }^{2}}\theta_T.
   \end{aligned}
\end{equation}

\begin{figure}[!t]
	\centering
	
	\renewcommand{\figurename}{Fig.}
	
	\begin{minipage}{0.48\linewidth}
		\centering
		\includegraphics[width=\linewidth]{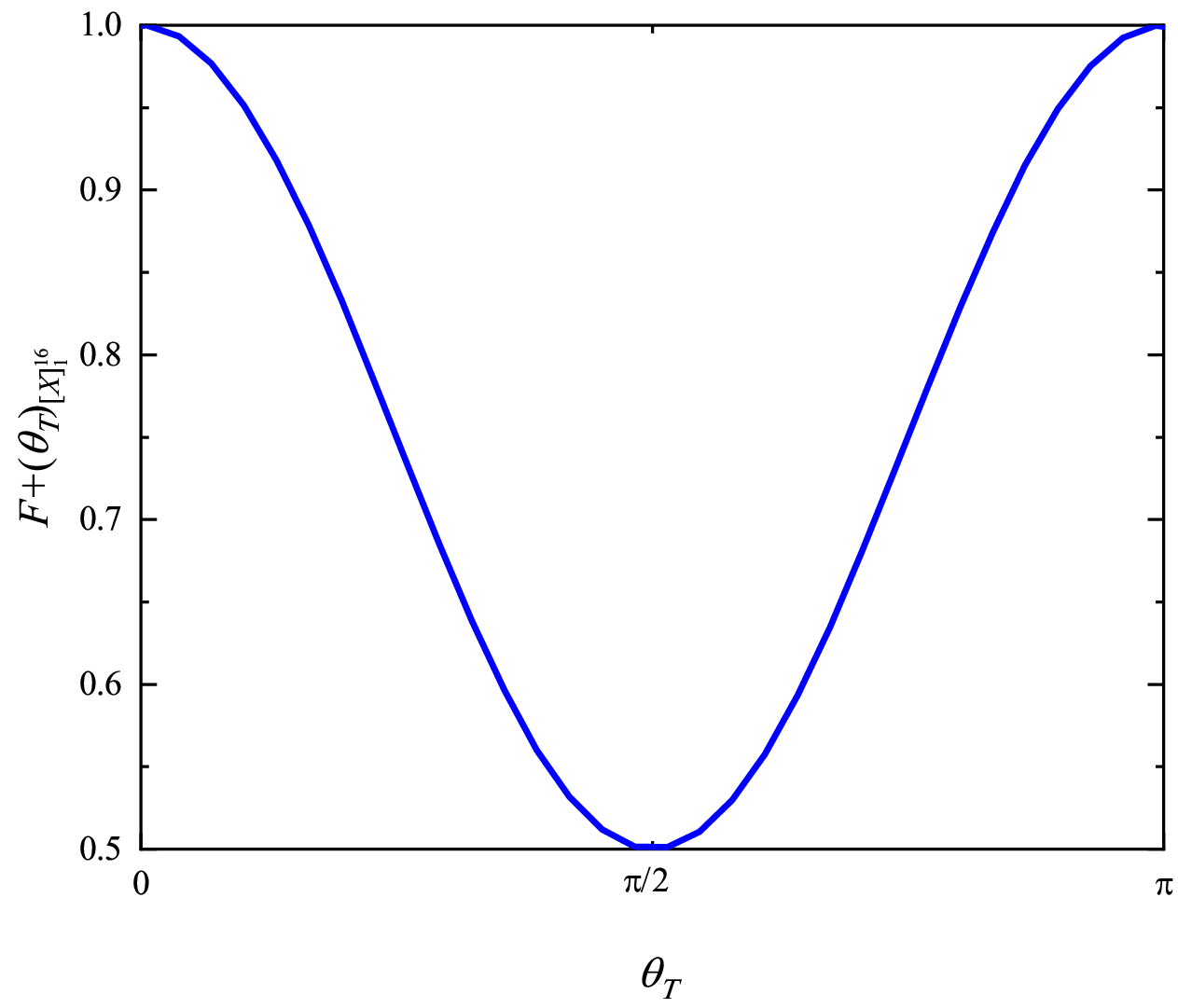}
		\subcaption{Plus mode}
	\end{minipage}
	\hfill
	\begin{minipage}{0.48\linewidth}
		\centering
		\includegraphics[width=\linewidth]{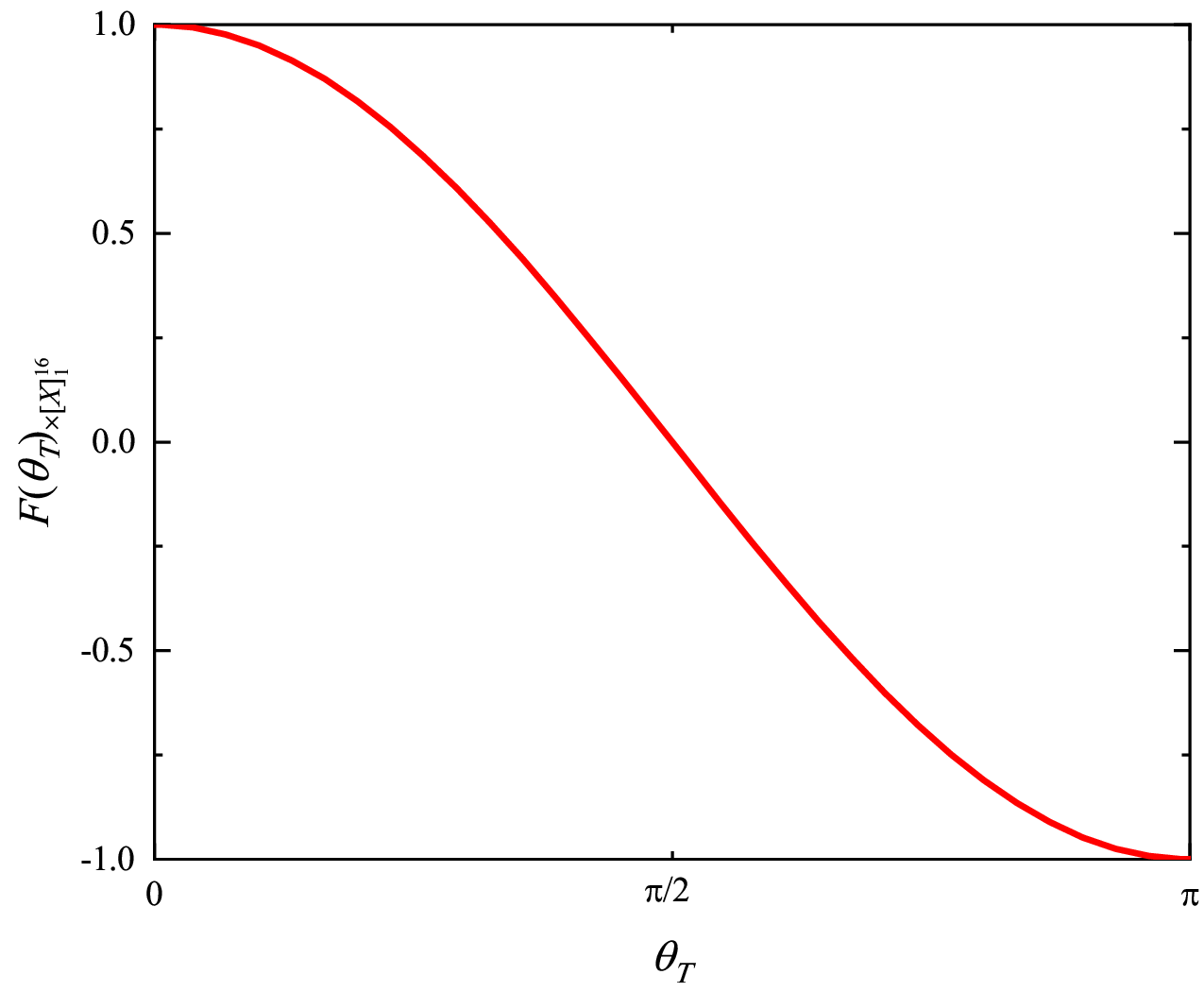}
		\subcaption{Cross mode}
	\end{minipage}
	
	\caption{\label{thetaT}
		The variation of the response function with respect to $\theta_L$.
		The left panel shows the plus modes, while the left panel shows the cross modes.
	}
\end{figure}

Specifically, for category 1, the tuple is given by $(n_a, n_b, n_c)= C_1\left(1,  0,  -1 \right){(iu)}^{2}$, where the constant $C_1$ is typically an integer.
As an example, for the modified second-generation Michelson combination $[X]_1^{16}$ with $C_1=16$, the response functions for the plus and cross modes are, respectively,
\begin{equation}
\begin{aligned}
	&\frac{4}{{{u}^{2}}}{{\left| F_{+}(\theta_T ) \right|}^{2}}_{\left[ X \right]_{1}^{16}}\to 384{{\left( \frac{{{\cos }^{2}}\theta_T +1}{2} \right)}^{2}},\\
	&\frac{4}{{{u}^{2}}}{{\left| F_{\times }(\theta_T ) \right|}^{2}}_{\left[ X \right]_{1}^{16}}{\mathop{\to }}\,384{{\cos }^{2}}\theta_T. \label{FPC_C1}
 \end{aligned}
\end{equation}
After integrating out $\phi_T$, the angular response function of  Michelson combination $[X]_1^{16}$ depends only on $\theta_T$.
The response functions in the plus modes and cross modes as a function of $\theta_T$ are shown in Fig.~\ref{thetaT}.

Both response functions attain the maxima at $\theta_T =0, \pi$, and reach the minima when $\theta_T =\frac{\pi }{2}$.
The above result also holds for the remainder combinations of category 1, such as $\left[ E \right]_{2}^{16}$ with $n=2$.
For the second and third categories, we have $(n_a, n_b, n_c)= C_2\left(-1, 1, 0\right){(iu)}^{2}$ and $(n_a, n_b, n_c)= C_3\left(0, 1, -1\right){(iu)}^{2}$. 
In particular, the combination $\left[ T \right]_{11}^{\text{16}}$ corresponds to $C_2=2$. 
It is straightforward to show that the resulting response functions are proportional to Eq.~\eqref{FPC_C1}, 
This is understood since the first three categories are related by a permutation operation on the detector plane.

For category 4, the tuple reads $(n_a, n_b, n_c)= C_4\left( 2,  -1, -1 \right){{(iu)}^{2}}$.
Taking $\left[ U \right]_{1}^{16}$ as an example, we have $C_4 = 2$ and
\begin{equation}
  \begin{aligned}
	&\frac{4}{{{u}^{2}}}{{\left| F_{+}(\theta_T ) \right|}^{2}}_{\left[ \text{U} \right]_{1}^{16}}\to 162{{\left( \frac{{{\cos }^{2}}\theta_T +1}{2} \right)}^{2}},\\
	&\frac{4}{{{u}^{2}}}{{\left| F_{\times }(\theta_T ) \right|}^{2}}_{\left[ \text{U} \right]_1^{16}}{\mathop{\to }}\,162{{\cos }^{2}}\theta_T.
	\end{aligned}	
\end{equation}
For category 5, such as $\left[ U \right]_{5}^{16}$ with $(n_a, n_b, n_c)= C_5\left( 4, -1, -3\right){{(iu)}^{2}}$ and $C_5=2$, one also has
\begin{equation}
 \begin{aligned}
	&\frac{4}{{{u}^{2}}}{{\left| F_{+}(\theta_T ) \right|}^{2}}_{\left[ \text{U} \right]_{5}^{16}}\to 78\left(\frac{ {{\cos }^{2}}\theta_T +1}{2} \right)^2,\\
	&\frac{4}{{{u}^{2}}}{{\left| F_{\times }(\theta_T ) \right|}^{2}}_{\left[ \text{U} \right]_{5}^{16}}{\mathop{\to }}\,78{{\cos }^{2}}\theta_T.
	\end{aligned}	
\end{equation}
Similarly, for category 6, such as $\left[ U \right]_{4}^{\text{16}}$ with $(n_a, n_b, n_c)= C_6\left( 8, -3, -5\right){{(iu)}^{2}}$ and $C_6=2$, we find
\begin{equation}
 \begin{aligned}
	&\frac{4}{{{u}^{2}}}{{\left| F_{+}(\theta_T ) \right|}^{2}}_{\left[ \text{U} \right]_{4}^{16}}\to 294{{\left( \frac{{{\cos }^{2}}\theta_T +1}{2} \right)}^{2}},\\
	&\frac{4}{{{u}^{2}}}{{\left| F_{\times }(\theta_T ) \right|}^{2}}_{\left[ \text{U} \right]_{4}^{16}}{\mathop{\to }}\,294{{\cos }^{2}}\theta_T.
  \end{aligned}
\end{equation}
Therefore, when the azimuthal angle is averaged for the first six categories of TDI combinations, the dependence on the zenithal angle is identical.
The response functions attain the maxima at $\theta_T =0, \pi$ and reach the minima at $\theta_T ={\pi }/{2}$.

Last but not least, for full-symmetric Sagnac TDI combinations, one must carry out a higher-order expansion as given in Eq.~\eqref{fullyappabcMod}.
At the low-frequency limit, the resultant response functions can be shown of the order $O(u^4)$.
Specifically, employing Eq.~\eqref{responfullyappabc}, the response function for the plus mode is found to be
\begin{equation}
	\frac{4}{{{u}^{2}}}{{\left| F_{+}(\theta_T ) \right|}^{2}}{{\left(\frac{6}{{{{\tilde{P}}}_{1}}{{(iu)}^{2}}}\right)}^{2}}=\frac{9}{32}{{\left({{\cos }^{2}}\theta_T +1\right)}^{2}}{{\sin }^{2}}\theta_T   
    =\frac{9}{32}\left( -{{\cos }^{6}}\theta_T -{{\cos }^{4}}\theta_T +{{\cos }^{2}}\theta_T +1 \right).
\end{equation}
One observes that the response function attains the maximum at $\theta_T =\arccos \frac{1}{\sqrt{3}}$, and minimum at $\theta_T =0, \pi$. 
Similarly, for the cross mode, we have
\begin{equation}
	\frac{4}{{{u}^{2}}}{{\left| F_{\times }(\theta_T ) \right|}^{2}}{{\left(\frac{6}{{{{\tilde{P}}}_{1}}{{(iu)}^{2}}}\right)}^{2}}=\frac{9}{8}{{\cos }^{2}}\theta_T {{\sin }^{2}}\theta_T =\frac{9}{32}{{\sin }^{2}}2\theta_T,
\end{equation}
which attains it maximum at $\theta_T ={\pi}/{4}$ and minimum at $\theta_T =0, \pi$.

\subsection{Beyond the low-frequency limit}\label{appSTran}

\begin{figure}[tbp]
	\centering
	\includegraphics[width=\linewidth]{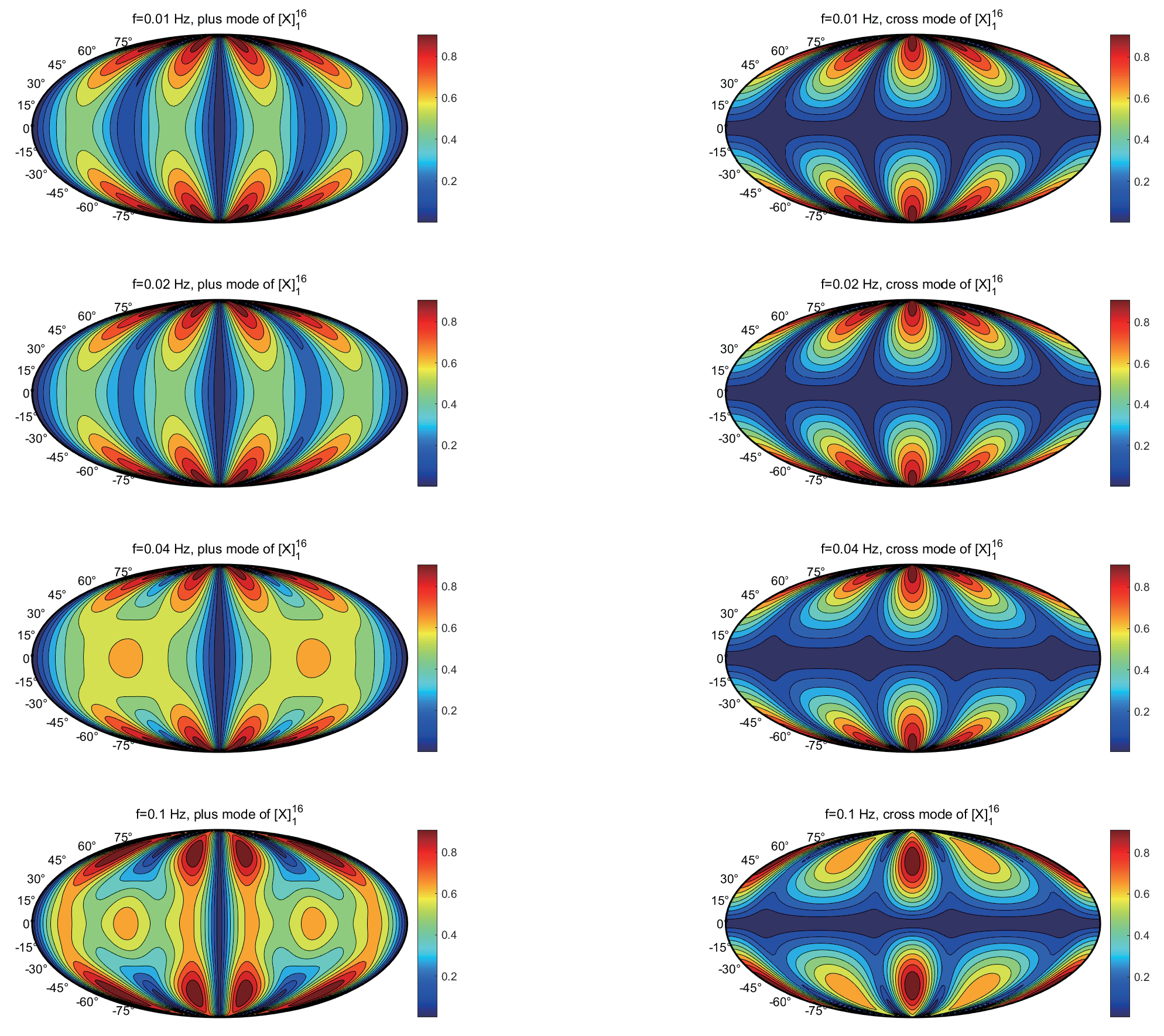}
	\caption{\label{figX1c} 
		The angular dependence of the response function of the TDI combination 
		$\left[ X\right]_{1}^{16}$ for the plus and cross modes, evaluated for different GW frequencies.}
\end{figure}

\begin{figure}[tbp]
	\centering
	\includegraphics[width=\linewidth]{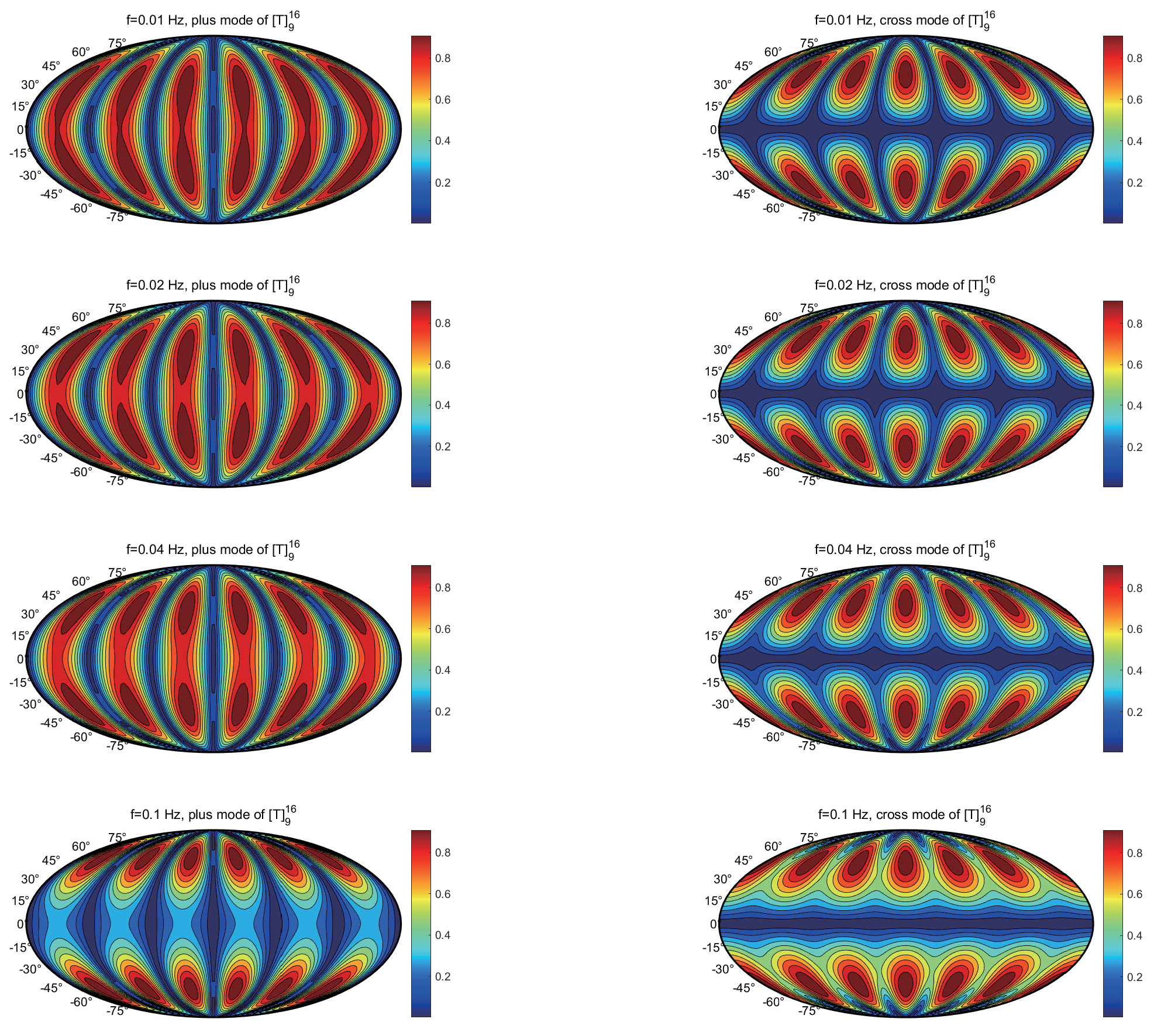}
	\caption{\label{figT9c} 
		The angular dependence of the response function of the TDI combination 
		$\left[ T\right]_{9}^{16}$ for the plus and cross modes, evaluated for different GW frequencies.
	}
\end{figure}

It is also interesting to ascertain the validity of the present findings, particularly regarding the low-frequency approximation. 
For more significant frequencies, the resulting angular distribution of the response function can also evaluated numerically.
The results are presented in Figs.~\ref{figX1c} and~\ref{figT9c}, by taking the Michelson combination $[X]_1^{16}$ and the fully-symmetric Sagnac combination $[T]_9^{16}$ as examples.
The angular distributions of the response functions are evaluated at frequencies $0.01$, $0.02$, $0.04$, and $0.1$ Hz.
By comparing the first row of Fig.~\ref{figX1c} against Fig.~\ref{LISAdetector}(a), the response function converges to the analytical results at minor frequencies.
However, as one deviates beyond the low-frequency region, specifically $f> 2\times 10^{-2}$ Hz, it is observed that the angular distribution gradually modifies.
As a result, the analytical results obtained in the present study cease to be a good approximation.
Nonetheless, our findings are pertinent in practice as the millihertz frequency region is the target band for the ongoing space-borne GW programs.

\section{The response function of the TDI combination in the barycentric frame}\label{section6}
For a fixed source direction $(\theta_B,\phi_B)$ in the barycentric frame, the apparent source direction in the detector frame varies with time.
This section will first analyze the angular response function of the $A, E, T$ channels in the barycentric frame, and then provide the response function of the optimal channel.
\subsection{The angle-dependent response function of the optimal channel in LISA detector}\label{section6.1}
The expressions for the Sagnac combinations $\alpha$, $\beta$, and $\gamma$ are as follows:
\begin{align}\label{Sagnac}
\alpha  =& {\eta _1} - {\eta _{1'}} + {D_3}{\eta _2} - {D_{2'1'}}{\eta _{2'}} + {D_{31}}{\eta _3} - {D_{2'}}{\eta _{3'}},\\\notag
\beta  =& {\eta _{\rm{2}}} - {\eta _{{\rm{2'}}}} + {D_{\rm{1}}}{\eta _{\rm{3}}} - {D_{{\rm{3'2'}}}}{\eta _{{\rm{3'}}}} + {D_{{\rm{12}}}}{\eta _{\rm{1}}} - {D_{{\rm{3'}}}}{\eta _{{\rm{1'}}}},\\\notag
\gamma  =& {\eta _{\rm{3}}} - {\eta _{{\rm{3'}}}} + {D_{\rm{2}}}{\eta _{\rm{1}}} - {D_{{\rm{1'3'}}}}{\eta _{{\rm{1'}}}} + {D_{{\rm{23}}}}{\eta _{\rm{2}}} - {D_{{\rm{1'}}}}{\eta _{{\rm{2'}}}}.
\end{align}
The Sagnac combinations are linearly combined to construct the optimal channels $A$, $E$, and $T$, one find
\begin{align}\label{aet}
A =& \frac{{\gamma  - \alpha }}{{\sqrt 2 }},\\\notag
E =& \frac{{\alpha  - 2\beta  + \gamma }}{{\sqrt 6 }},\\\notag
T =& \frac{{\alpha  + \beta  + \gamma }}{{\sqrt {\rm{3}} }}.
\end{align}
Therefore, the expression for channel $A$ is
\begin{align}\label{a}
A = \frac{1}{\sqrt 2 }\left[ {\left( { - {\rm{1 + }}{D_{\rm{2}}}} \right){\eta _1}{\rm{ + }}\left( {{\rm{1}} - {D_{{\rm{1'3'}}}}} \right){\eta _{1'}} + \left( { - {D_3}{\rm{ + }}{D_{{\rm{23}}}}} \right){\eta _{\rm{2}}}{\rm{ + }}\left( { - {D_{{\rm{1'}}}}{\rm{ + }}{D_{2'1'}}} \right){\eta _{2'}}{\rm{ + }}\left( {{\rm{1}} - {D_{31}}} \right){\eta _{\rm{3}}}{\rm{ + }}\left( { - {\rm{1 + }}{D_{2'}}} \right){\eta _{3'}}} \right].
\end{align}
The polynomial coefficients of its time-delay operator are
\begin{align}\label{apoco}
{P_1} =& \frac{1}{\sqrt 2 }\left( - 1 +{D_2} \right),{P_2} = \frac{1}{\sqrt 2 }\left(  - {D_3}+D_{23} \right),{P_3} = \frac{1}{\sqrt 2 }\left( 1 - D_{31} \right),\\\notag
{P_{1'}} =& \frac{1}{\sqrt 2 }\left( 1 - D_{1'3'} \right),{P_{2'}} = \frac{1}{\sqrt 2 }\left( - D_{1'} + D_{2'1'} \right),P_{3'} = \frac{1}{\sqrt 2 }\left(  - 1 + D_{2'} \right).
\end{align}
The Fourier transform of the coefficients is
\begin{align}\label{apocoFou}
{{\tilde P}_{\rm{1}}} =& \frac{{\rm{1}}}{{\sqrt 2 }}\left( { - {\rm{1 + }}{e^{iu}}} \right),{{\tilde P}_{\rm{2}}} = \frac{{\rm{1}}}{{\sqrt 2 }}\left( { - {e^{iu}}+ {e^{i2u}}} \right),{{\tilde P}_{\rm{3}}} = \frac{{\rm{1}}}{{\sqrt 2 }}\left( {{\rm{1}} - {e^{i2u}}} \right),\\\notag
{{\tilde P}_{{\rm{1'}}}} =& \frac{{\rm{1}}}{{\sqrt 2 }}\left( {{\rm{1}} - {e^{i2u}}} \right),{{\tilde P}_{{\rm{2'}}}} = \frac{{\rm{1}}}{{\sqrt 2 }}\left( { - {e^{iu}}+ {e^{i2u}}} \right),{{\tilde P}_{{\rm{3'}}}} = \frac{{\rm{1}}}{{\sqrt 2 }}\left( { - {\rm{1 + }}{e^{iu}}} \right).
\end{align}
We have
\begin{align}\label{apocoFoup}
{{\tilde P}_{\rm{1}}} + {{\tilde P}_{{\rm{2'}}}} =  - \frac{{\rm{1}}}{{\sqrt 2 }}\left( {{\rm{1}} - {e^{i2u}}} \right),\\\notag
{{\tilde P}_{\rm{2}}} + {{\tilde P}_{{\rm{3'}}}} =  - \frac{{\rm{1}}}{{\sqrt 2 }}\left( {{\rm{1}} - {e^{i2u}}} \right),\\\notag
{{\tilde P}_{\rm{3}}} + {{\tilde P}_{{\rm{1'}}}} = \frac{2}{{\sqrt 2 }}\left( {{\rm{1}} - {e^{i2u}}} \right).
\end{align}
Substituting Eqs.~\eqref{apocoFoup} into Eqs.~\eqref{responappabc}, the response function of the plus mode and cross mode in channel $A$ under the detector's reference frame are
\begin{align}\label{aresponse}
\frac{4}{{{u^2}}}{\left| {F_ + ^{(A)}} \right|^2}\mathop  \to \limits_{u <  < 1}& \frac{9}{2}{\sin ^2}u{\left(\frac{ {{{\cos }^2}\theta_L  + 1} }{2}\right)^2}{\left( { - \cos 2\phi_L  + \sqrt 3 \sin 2\phi_L } \right)^2},\\\notag
\frac{4}{{{u^2}}}{\left| {F_ \times ^{(A)}} \right|^2}\mathop  \to \limits_{u <  < 1}& \frac{9}{2}{\sin ^2}u\cos^2\theta_L {\left( {\sqrt 3 \cos 2\phi_L  + \sin 2\phi_L } \right)^2}.
\end{align}

\begin{figure}[tbp]
	\centering
	\includegraphics[width=\linewidth]{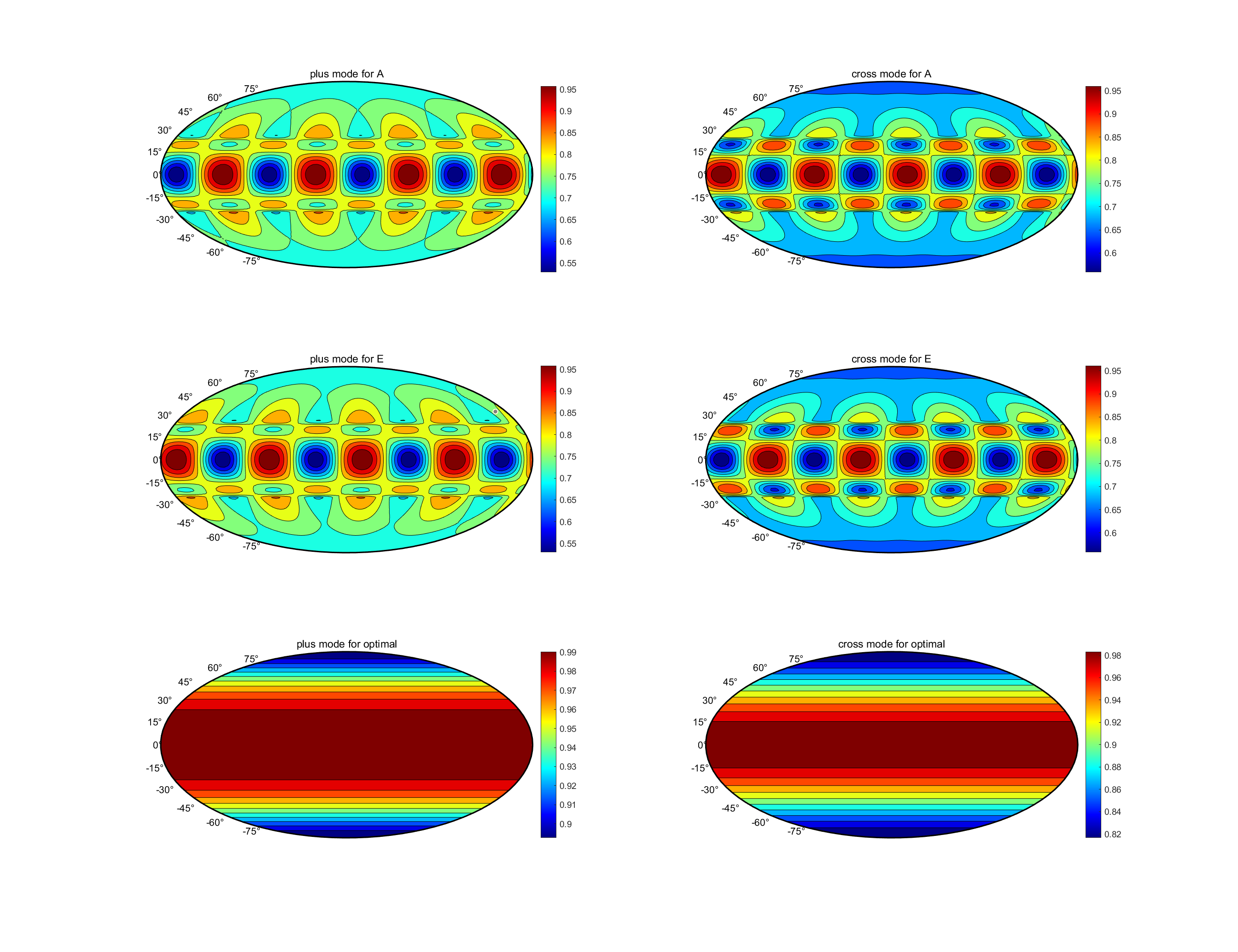}
	\caption{\label{optimal} 
		The angular dependence of the response function of the channels $A$, $E$, and optimal channel 
		for the plus and cross modes, where the plotting is done with normalization.
	}
\end{figure}

Using the relationship between the barycentric frame and the LISA detector frame, Eqs.~\eqref{LISAbary}, and substituting it into Eqs.~\eqref{aresponse}, the angle dependence of channel $A$ in the barycentric frame is shown in the first row of Fig.~\ref{optimal}. 
It can be observed that the angular response function of channel $A$ is dependent on angle $\phi_B$.

The expression for channel $E$ is
\begin{align}
E=\frac{1}{\sqrt 6 }\left[ \begin{array}{l}
\left( {1 + {D_{\rm{2}}} - {\rm{2}}{D_{{\rm{12}}}}} \right){\eta _1} + \left( { - 1 + 2{D_{{\rm{3'}}}} - {D_{{\rm{1'3'}}}}} \right){\eta _{1'}} + \left( { - 2 + {D_3} + {D_{{\rm{23}}}}} \right){\eta _2}\\
 + \left( {2 - {D_{{\rm{1'}}}} - {D_{2'1'}}} \right){\eta _{2'}} + \left( {1 - 2{D_{\rm{1}}} + {D_{31}}} \right){\eta _3} + \left( { - 1 - {D_{2'}} + 2{D_{{\rm{3'2'}}}}} \right){\eta _{3'}}
\end{array} \right].
\end{align}
The polynomial coefficients of its time-delay operator are
\begin{align}
{P_1} =& \frac{{\rm{1}}}{{\sqrt 6 }}\left( {1 + {D_{\rm{2}}} - {\rm{2}}{D_{{\rm{12}}}}} \right),{P_2} = \frac{{\rm{1}}}{{\sqrt 6 }}\left( { - 2 + {D_3} + {D_{{\rm{23}}}}} \right),{P_3} = \frac{{\rm{1}}}{{\sqrt 6 }}\left( {1 - 2{D_{\rm{1}}} + {D_{31}}} \right),\\\notag
{P_{1'}} =& \frac{{\rm{1}}}{{\sqrt 6 }}\left( { - 1 + 2{D_{{\rm{3'}}}} - {D_{{\rm{1'3'}}}}} \right),{P_{2'}} = \frac{{\rm{1}}}{{\sqrt 6 }}\left( {2 - {D_{{\rm{1'}}}} - {D_{2'1'}}} \right),{P_{3'}} = \frac{{\rm{1}}}{{\sqrt 6 }}\left( { - 1 - {D_{2'}} + 2{D_{{\rm{3'2'}}}}} \right).
\end{align}
The Fourier transform of its coefficients are
\begin{align}\label{epocoFou}
{{\tilde P}_1} =& \frac{{\rm{1}}}{{\sqrt 6 }}\left( {1 + {e^{iu}} - {\rm{2}}{e^{i2u}}} \right),{{\tilde P}_2} = \frac{{\rm{1}}}{{\sqrt 6 }}\left( { - 2 + {e^{iu}} + {e^{i2u}}} \right),{{\tilde P}_3} = \frac{{\rm{1}}}{{\sqrt 6 }}\left( {1 - 2{e^{iu}} + {e^{i2u}}} \right),\\\notag
{{\tilde P}_{1'}} =& \frac{{\rm{1}}}{{\sqrt 6 }}\left( { - 1 + 2{e^{iu}} - {e^{i2u}}} \right),{{\tilde P}_{2'}} = \frac{{\rm{1}}}{{\sqrt 6 }}\left( {2 - {e^{iu}} - {e^{i2u}}} \right),{{\tilde P}_{3'}} = \frac{{\rm{1}}}{{\sqrt 6 }}\left( { - 1 - {e^{iu}} + 2{e^{i2u}}} \right).
\end{align}
It can be obtained as
\begin{align}\label{epocoFoup}
{{\tilde P}_1} + {{\tilde P}_{2'}} =& \frac{3}{{\sqrt 6 }}\left( {1 - {e^{i2u}}} \right),\\\notag
{{\tilde P}_2} + {{\tilde P}_{3'}} =&  - \frac{3}{{\sqrt 6 }}\left( {1 - {e^{i2u}}} \right),\\\notag
{{\tilde P}_3} + {{\tilde P}_{1'}} =& 0.
\end{align}
Substituting Eqs.~\eqref{epocoFoup} into Eqs.~\eqref{responappabc}, the response function of the plus mode and cross mode in channel $E$ under the detector's reference frame are
\begin{align}\label{eresponse}
\frac{4}{{{u^2}}}{\left| {F_ + ^{(E)}} \right|^2}\mathop  \to \limits_{u <  < 1} &\frac{9}{2}{\sin ^2}u{\left(\frac{ {{{\cos }^2}\theta_L  + 1}}{2} \right)^2}{\left( {\sqrt 3 \cos 2\phi_L  + \sin 2\phi_L } \right)^2},\\\notag
\frac{4}{{{u^2}}}{\left| {F_ \times ^{(E)}} \right|^2}\mathop  \to \limits_{u <  < 1}& \frac{9}{2}{\sin ^2}u\cos^2\theta_L {\left( { - \cos 2\phi_L  + \sqrt 3 \sin 2\phi_L } \right)^2}.
\end{align}
Using the relationship between the barycentric frame and the LISA detector frame, and substituting Eqs.~\eqref{LISAbary} into the Eqs.~\eqref{eresponse}, the angle dependence of channel $E$ in the barycentric frame is shown in the second row of Fig.~\ref{optimal}.
It can be observed that channel $E$ is dependent on angle $\phi_B$.

For channel $T$, its expression is
\begin{align}
T = \frac{{\rm{1}}}{{\sqrt {\rm{3}} }}\left[ \begin{array}{l}
\left( {1{\rm{ + }}{D_{\rm{2}}}{\rm{ + }}{D_{{\rm{12}}}}} \right){\eta _1} - \left( {1 + {D_{{\rm{3'}}}} + {D_{{\rm{1'3'}}}}} \right){\eta _{1'}}\\
 + \left( {1 + {D_3} + {D_{{\rm{23}}}}} \right){\eta _2} - \left( {1 + {D_{{\rm{1'}}}} + {D_{2'1'}}} \right){\eta _{2'}}\\
 + \left( {1 + {D_{\rm{1}}} + {D_{31}}} \right){\eta _3} - \left( {1 + {D_{2'}} + {D_{{\rm{3'2'}}}}} \right){\eta _{3'}}
\end{array} \right],
\end{align}
The polynomial coefficients of its time-delay operator are
\begin{align}
{P_1} =& \frac{{\rm{1}}}{{\sqrt {\rm{3}} }}\left( {1 + {D_{\rm{2}}} + {D_{{\rm{12}}}}} \right),{P_2} = \frac{{\rm{1}}}{{\sqrt {\rm{3}} }}\left( {1 + {D_3} + {D_{{\rm{23}}}}} \right),{P_3} = \frac{{\rm{1}}}{{\sqrt {\rm{3}} }}\left( {1 + {D_{\rm{1}}} + {D_{31}}} \right),\\\notag
{P_{1'}} =&  - \frac{{\rm{1}}}{{\sqrt {\rm{3}} }}\left( {1 + {D_{{\rm{3'}}}} + {D_{{\rm{1'3'}}}}} \right),{P_{2'}} =  - \frac{{\rm{1}}}{{\sqrt {\rm{3}} }}\left( {1 + {D_{{\rm{1'}}}} + {D_{2'1'}}} \right),{P_{3'}} =  - \frac{{\rm{1}}}{{\sqrt {\rm{3}} }}\left( {1 + {D_{2'}} + {D_{{\rm{3'2'}}}}} \right).
\end{align}
The Fourier transform of the coefficients for channel $T$ is
\begin{align}\label{tpocoFou}
{{\tilde P}_1} =& {{\tilde P}_2} = {{\tilde P}_3} = \frac{{\rm{1}}}{{\sqrt {\rm{3}} }}\left( {1 + {e^{iu}} + {e^{i2u}}} \right),\\\notag
{{\tilde P}_{1'}} =& {{\tilde P}_{2'}} = {{\tilde P}_{3'}} =  - \frac{{\rm{1}}}{{\sqrt {\rm{3}} }}\left( {1 + {e^{iu}} + {e^{i2u}}} \right).
\end{align}
We have
\begin{align}\label{tpocoFoup}
{{\tilde P}_1} + {{\tilde P}_{2'}} = 0,\\\notag
{{\tilde P}_2} + {{\tilde P}_{3'}} = 0,\\\notag
{{\tilde P}_3} + {{\tilde P}_{1'}} = 0.
\end{align}
Substituting Eqs.~\eqref{tpocoFoup} into Eqs.~\eqref{responappabc}, it is found that the response functions of both the plus mode and the cross mode in channel $T$ are zero in the low-frequency approximation.
Based on the literature~\citep{optimal-SNR-2003}, the optimization problem can be reduced to an eigenvalue problem, where the eigenvalue is the square of the SNR.
Substituting the polynomial coefficients from Eqs.~\eqref{apocoFou},\eqref{epocoFou}, and \eqref{tpocoFou} into Eqs.~\eqref{tmnoisepsd} and \eqref{opnoisepsd}, 
the noise PSD for the $A$, $E$, and $T$ channels are obtained as:
\begin{align}\label{AETNOISE}
{N_A}(u) =& {N_E}(u) = {\sin ^2}\frac{u}{2}\left[ {16\left( {3 + 2\cos u + \cos 2u} \right) \times \frac{{{L^2}s_a^2}}{{{u^2}{c^4}}} + 8\left( {2 + \cos u} \right) \times \frac{{{u^2}s_x^2}}{{{L^2}}}} \right],\\\notag
{N_T}(u) =& \left( {4 - 4\cos 3u} \right) \times \frac{{{L^2}s_a^2}}{{{u^2}{c^4}}} + 2{\left( {1 + 2\cos u} \right)^2} \times \frac{{{u^2}s_x^2}}{{{L^2}}}.
\end{align}

Therefore, the SNR of the optimal combination depends on
\begin{align}\label{plusSNR}
{\rm{SNR}}_ + ^2 \propto& \frac{4}{{{u^2}}}{\left| {F_ + ^{(A)}} \right|^2} + \frac{4}{{{u^2}}}{\left| {F_ + ^{(E)}} \right|^2} + \frac{4}{{{u^2}}}{\left| {F_ + ^{(T)}} \right|^2}\\\notag
 =& 18{\sin ^2}u{\left( \frac{{{{\cos }^2}\theta_L  + 1}}{2} \right)^2},
\end{align}
and
\begin{align}\label{crossSNR}
{\rm{SNR}}_ \times ^2 \propto& \frac{4}{{{u^2}}}{\left| {F_ \times ^{(A)}} \right|^2} + \frac{4}{{{u^2}}}{\left| {F_ \times ^{(E)}} \right|^2} + \frac{4}{{{u^2}}}{\left| {F_ \times ^{(T)}} \right|^2}\\\notag
 =& 18{\sin ^2}u\cos^2\theta_L. 
\end{align}
Using the relationship between the detector coordinate system and the solar barycentric frame, the plus mode is integrated as follows
\begin{align}\label{INTERplusSNR}
{\rm{SNR}}_ + ^2(\cos {\theta _L}) \propto \frac{1}{{{T_ \odot }}}\int_0^{{T_ \odot }} {{{\left(\frac{{{{\cos }^2}{\theta_L} + 1}}{2} \right)}^2}} dt.
\end{align}
By integrating the SNR of the cross mode, we can obtain
\begin{align}\label{INTERcrossSNR}
{\rm{SNR}}_ \times ^2(\cos {\theta_L}) \propto \frac{1}{{{T_ \odot }}}\int_0^{{T_ \odot }} {{{\cos }^2}{\theta_L}} dt,
\end{align}
where $T_\odot $ is the integration time.
By taking one year of integration, we can obtain
\begin{align}\label{INTERthetaBplusSNR}
{\rm{SNR}}_ + ^2(\cos {\theta _B}) \propto& \frac{1}{{2\pi }}\frac{1}{2}\int_0^{2\pi } {\left\{ {{{\left[ {\frac{1}{2}\cos {\theta _B} - \frac{{\sqrt 3 }}{2}\sin {\theta _B}\sin \left( {{\phi _B} - \omega t} \right)} \right]}^2} + 1} \right\}}^2 dt\\\notag
 \propto& \frac{1}{{4096}}\left( { - 1841 + 204\cos 2{\theta_B} + 37\cos 4{\theta _B}} \right),
\end{align}
and
\begin{align}\label{INTERthetaBcrossSNR}
{\rm{SNR}}_ \times ^2(\cos {\theta _B})\propto& \frac{1}{{2\pi }}{\int_0^{2\pi } {\left[ {\frac{1}{2}\cos {\theta _B} - \frac{{\sqrt 3 }}{2}\sin {\theta _B}\sin \left( {{\phi _B} - \omega t} \right)} \right]} ^2}dt\\\notag
 \propto& - \frac{1}{{16}}\left( { - 5 + \cos 2{\theta _B}} \right).
\end{align}
Therefore, it can be observed that although the SNR of the $A$, $E$, and $T$ channels depend on the angle $\phi_B$, the SNR of the optimal channel does not depend on the angle $\phi_B$, as shown in the third row of Fig.~\ref{optimal}.
The following introduces the angle dependence of the response function for a single TDI combination.

\subsection{The angular response function of the single TDI combination in LISA detector. }\label{section5.2}
\begin{figure}[tbp]
	\centering
	\includegraphics[width=\linewidth]{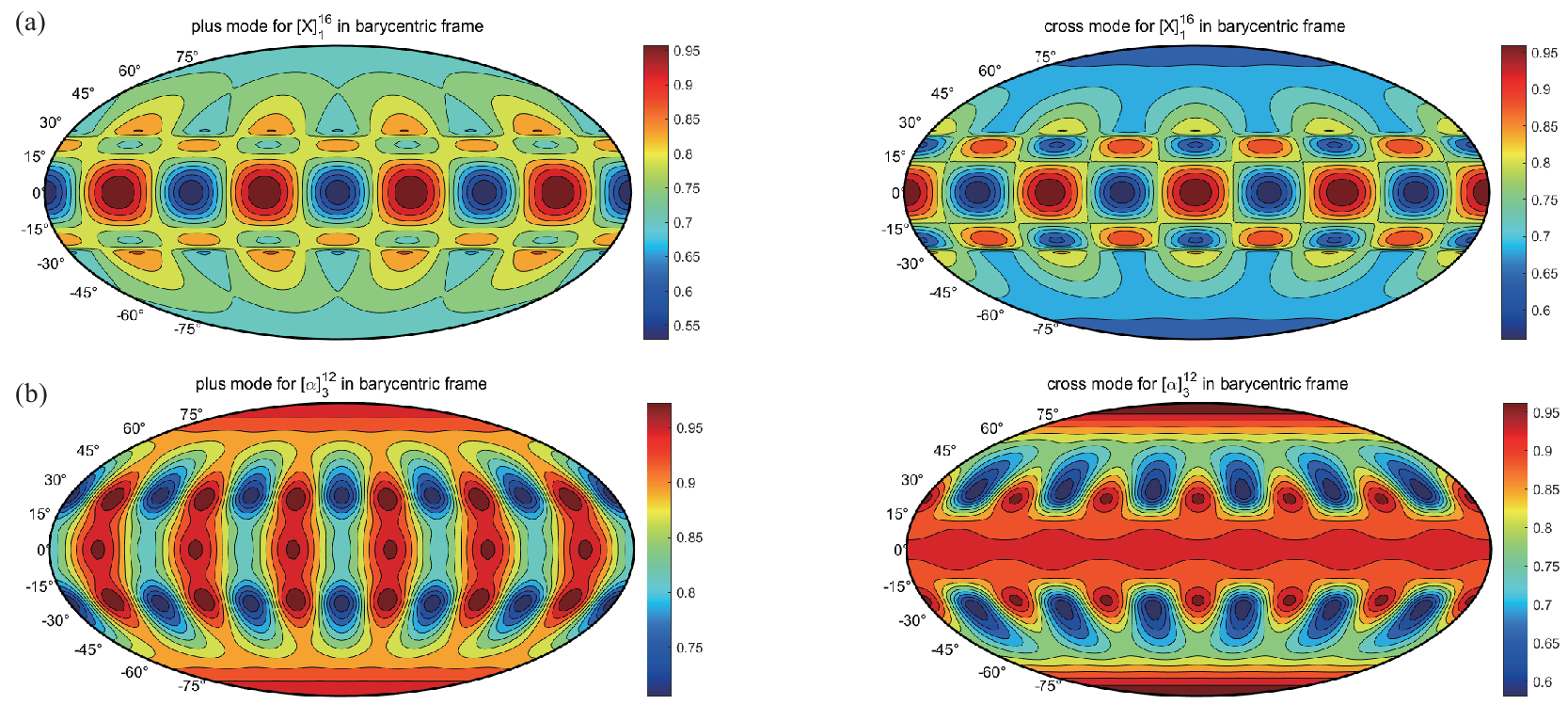}
	\caption{\label{LISA_baryresponse}
		(a) The angular dependence of the response function of the combination $[X]_1^{16}$ for the plus and cross modes, 
		(b) the angular dependence of the response function of the combination $[\alpha]_3^{12}$ for the plus and cross modes, 
		where the plotting is done with normalization.
	}
\end{figure}

Taking the second-generation Michelson combination as an example, according to Eq.~\eqref{X1four}, we can derive
\begin{align}\label{X1fourp}
{{\tilde P}_1} + {{\tilde P}_{2'}} =& (1 + {e^{iu}})\left( {1 - {e^{2iu}} - {e^{4iu}} + {e^{6iu}}} \right),\\\notag
{{\tilde P}_2} + {{\tilde P}_{3'}} =& 0,\\\notag
{{\tilde P}_3} + {{\tilde P}_{1'}} =&  - (1 + {e^{iu}})\left( {1 - {e^{2iu}} - {e^{4iu}} + {e^{6iu}}} \right).
\end{align}
Substituting Eq.~\eqref{X1fourp} into Eq.~\eqref{responappabc}, one obtain:
\begin{align}\label{RESOPNSEX1fourp}
\frac{4}{{{u^2}}}{\left| {F_ + ^{(\left[ X \right]_1^{16})}} \right|^2}\mathop  \to \limits_{u <  < 1}& 48{\cos ^2}u{\csc ^2}\frac{u}{2}{\sin ^6}u{\left( {{{\cos }^2}{\theta _L} + 1} \right)^2}{\sin ^2}2{\phi _L},\\\notag
\frac{4}{{{u^2}}}{\left| {F_ \times ^{(\left[ X \right]_1^{16})}} \right|^2}\mathop  \to \limits_{u <  < 1}& 192{\cos ^2}u{\csc ^2}\frac{u}{2}{\sin ^6}u\cos^2{\theta _L}{\cos ^2}2{\phi _L}.
\end{align}
Using the relationship between the barycentric frame and the LISA detector frame, and substituting Eqs.~\eqref{LISAbary} into the Eqs.~\eqref{RESOPNSEX1fourp}, it can be observed that the response function of the Michelson combination $[X]_1^{16}$ is related to angle $\phi_B$, as shown in Fig.~\ref{LISA_baryresponse}(a).

Taking the fully symmetric Sagnac combination $[\alpha]_3^{12}$ as an example, the Fourier transform of the polynomial coefficients of its time-delay operator is
\begin{align}\label{alpha3coff}
{{\tilde P}_1} =& 1 - {e^{iu}},\\\notag
{{\tilde P}_2} =& 1 - {e^{iu}},\\\notag
{{\tilde P}_3} =& 1 - {e^{iu}},\\\notag
{{\tilde P}_{1'}} =&  - \left( {1 - {e^{iu}}} \right),\\\notag
{{\tilde P}_{2'}} =&  - \left( {1 - {e^{iu}}} \right),\\\notag
{{\tilde P}_{3'}} =&  - \left( {1 - {e^{iu}}} \right).
\end{align}
Substituting Eq.~\eqref{alpha3coff} into Eqs.~\eqref{responfullyappabc}, the response functions of the plus mode and cross mode for $[\alpha]_3^{12}$ can be expressed as:
\begin{align}\label{RESOPNSEalpha3fourp}
\frac{4}{{{u^2}}}{\left| {F_ + ^{(\left[ \alpha  \right]_{\rm{3}}^{{\rm{12}}})}} \right|^2}\mathop  \to \limits_{u <  < 1}&2(1 - \cos u) {\left( {\frac{{{{(iu)}^2}}}{6}} \right)^2}\frac{9}{{16}}{({\cos ^2}{\theta _L} + 1)^2}{\sin ^2}{\theta _L}{\sin ^2}3{\phi _L},\\\notag
\frac{4}{{{u^2}}}{\left| {F_ \times ^{\left[ \alpha  \right]_{\rm{3}}^{{\rm{12}}}}} \right|^2}\mathop  \to \limits_{u <  < 1}& \frac{1}{2}(1 - \cos u){\left( {\frac{{{{(iu)}^2}}}{6}} \right)^2}{\cos ^2}{\theta _L}{\sin ^2}{\theta _L}{\cos ^2}{\phi _L}{\left( {1 + 2\cos 2{\phi _L}} \right)^2}.
\end{align}
Using the relationship between the barycentric frame and the LISA detector frame, and substituting Eqs.~\eqref{LISAbary} into the Eqs.~\eqref{RESOPNSEalpha3fourp}. 
The response function of the fully symmetric Sagnac TDI combination $[\alpha]_3^{12}$ also depends on angle $\phi_B$, as shown in Fig.~\ref{LISA_baryresponse}(b).

\subsection{The angle-dependent response function of the optimal channel for TianQin detector}\label{section5.2}
\begin{figure}[tbp]
	\centering
	\includegraphics[width=\linewidth]{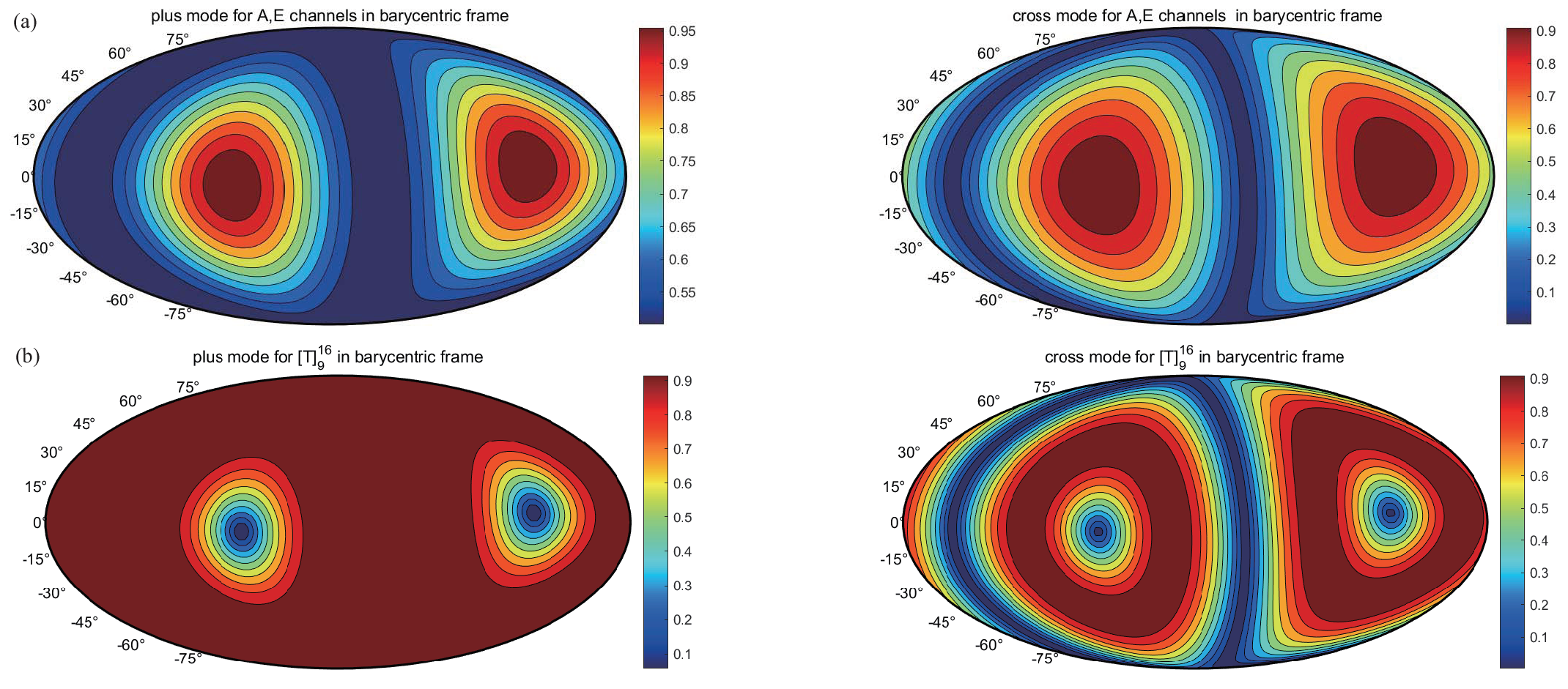}
	\caption{\label{TQ_baryresponse}
		(a) The angular dependence of the response function of the combination $A, E$ and optimal channel for the plus and cross modes in TianQin detector. 
		(b) The angular dependence of the response function of fully symmetric Sagnac TDI combination for the plus and cross modes in TianQin detector.
	}
\end{figure}

Substituting Eqs.~\eqref{apocoFoup} and \eqref{epocoFoup} into Eqs.~\eqref{responappabc}, and averaging $\phi_T$, the response functions for the plus mode and cross mode in the $A$ and $E$ channels in the TianQin detector's reference frame are
\begin{align}\label{TQae}
\frac{4}{{{u^2}}}{\left| {F_ + ^{(A)}{(}{\theta _T}{)}} \right|^2} =& \frac{4}{{{u^2}}}{\left| {F_ + ^{(E)}{\rm{(}}{\theta _T}{\rm{)}}} \right|^2}\mathop  \to \limits_{u <  < 1} 9{\sin ^2}u{\left( {\frac{{{{\cos }^2}{\theta _T} + 1}}{{\rm{2}}}} \right)^2},\\\notag
\frac{4}{{{u^2}}}{\left| {F_ \times ^{(A)}{\rm{(}}{\theta _T}{\rm{)}}} \right|^2} =& \frac{4}{{{u^2}}}{\left| {F_ + ^{(E)}{\rm{(}}{\theta _T}{\rm{)}}} \right|^2}\mathop  \to \limits_{u <  < 1} 9{\sin ^2}u\cos^2{\theta _T}.
\end{align}
Substituting Eqs.~\eqref{tpocoFoup} into Eqs.~\eqref{responappabc}, it is found that the response functions of both the plus mode and the cross mode in channel $T$ are zero in the low-frequency approximation.
Thus, the SNR of the optimal channel in the TianQin detector depends on:
\begin{align}\label{SNRtianplus}
{\rm{SNR}}_ + ^2 \propto& \frac{4}{{{u^2}}}{\left| {F_ + ^{(A)}({\theta _T})} \right|^2} + \frac{4}{{{u^2}}}{\left| {F_ + ^{(E)}({\theta _T})} \right|^2} + \frac{4}{{{u^2}}}{\left| {F_ + ^{(T)}({\theta _T})} \right|^2}\\\notag
 =& 18{\sin ^2}u{\left( {\frac{{{{\cos }^2}{\theta _T} + 1}}{2}} \right)^2},
\end{align}
and
\begin{align}\label{SNRtiancross}
{\rm{SNR}}_ \times ^2 \propto& \frac{4}{{{u^2}}}{\left| {F_ \times ^{(A)}{\rm{(}}{\theta _T}{\rm{)}}} \right|^2} + \frac{4}{{{u^2}}}{\left| {F_ \times ^{(E)}{\rm{(}}{\theta _T}{\rm{)}}} \right|^2} + \frac{4}{{{u^2}}}{\left| {F_ \times ^{(T)}{\rm{(}}{\theta _T}{\rm{)}}} \right|^2}\\\notag
 =& {\rm{18}}{\sin ^2}u{\cos^2}{\theta _T}.
\end{align}
Substituting Eq.~\eqref{TQbary} into Eq.~\eqref{SNRtianplus}, and integrating over the source trajectory of the plus mode:
\begin{align}\label{SNRtianplusint}
{\rm {SNR}}_ + ^2(\cos {\theta _B}) \propto \frac{{\rm{1}}}{{\rm{4}}}{\left\{ {{{\left[ {\cos {\theta _B}\cos {\theta _J} + \cos \left( {{\phi _B} - {\phi _J}} \right)\sin {\theta _B}\sin {\theta _J}} \right]}^2} + 1} \right\}^{\rm{2}}}.
\end{align}
Substituting Eq.~\eqref{TQbary} into Eq.~\eqref{SNRtiancross}, and integrating over the source trajectory of the aross mode:
\begin{align}\label{SNRtiancrossint}
{\rm {SNR}}_ \times ^2(\cos {\theta _B}) \propto {\left[ {\cos {\theta _B}\cos {\theta _J} + \cos \left( {{\phi _B} - {\phi _J}} \right)\sin {\theta _B}\sin {\theta _J}} \right]^2}.
\end{align}
Since the TianQin detector plane always faces the reference source RX J0806.3+1527, the SNR of the optimal combination depends on the angle between the detected source and the reference wave source, as shown in Fig.~\ref{TQ_baryresponse}(a). 
Similarly, the response functions for other TDI combinations are also analogous.
For the fully symmetric Sagnac TDI combination, the response function with respect to $\phi_B$ and $\theta_B$ is shown in  Fig.~\ref{TQ_baryresponse}(b).

\section{Concluding remarks}\label{section7}

In this work, we focus on the dependence of the space-borne GW detector's response on the orientation of the source for different TDI combinations.
Since space gravitational wave detection mainly focuses on the millihertz frequency band, we provides angular dependence graphs for different TDI combinations under low-frequency approximations. 
The same angular dependence graph can be used to analyze the response intensity of the same TDI combination at different $\theta_D$ and $\phi_D$.
Analytical results are derived and explored for the forty-five geometric TDI combinations up to sixteen links.
We observed that the resultant angular dependence can be classified into seven distinct categories. 
To analyze the SNR for different TDI combinations, the signal needs to be integrated over the source trajectory.
For the LISA detector, the trajectory period is one year; for sources that can be identified with a high SNR in a short time, short-duration data can be selected, where $\theta_L$ and $\phi_L$ remain nearly unchanged under the detector frame.
For the TianQin detector, the angle $\theta_T$ is fixed and the period of $\phi_T$ is 3.65 days; averaging over $\phi_T$ corresponds to integrating over the source trajectory for analysis, which allows us to derive an analytical expression for the response function that depends solely on $\theta_T$.
Because a GW source’s angular location and central frequency might be unknown prior to the detection, the solid-angle and frequency averages are often performed when analyzing the sensitivity function of a given TDI combination. 
The present study, however, aims to explore the details of such dependence. 
This specific aspect is physically pertinent because once the initial detection is achieved, the obtained source’s location and frequency can be fed back to furnish a refined TDI combination tailored to the specific GW source. 
Since the TDI technique is a post-processing algorithm, such a procedure is feasible in practice.

%
%

\appendix

\setcounter{table}{0}
\renewcommand{\thetable}{A\arabic{table}}

\section{A summary of three relevant combinations of TDI coefficients for all forty-five geometric TDI solutions}\label{threeTerms}

In this appendix, we evaluate the three terms $\tilde{P}_i+\tilde{P}_{(i+1)'}$ for all forty-five geometric TDI solutions up to sixteen links.
The results are presented in Tabs.~\ref{12linkexpression}-\ref{16linkexpression17}. 
We note that some of the sixteen-link solutions denoted by $*$ in Tabs.~\ref{16linkexpression}-\ref{16linkexpression17} correspond to the combinations whose laser noise residuals possess the first-order contributions of the form $\dot{L}_i(L_j-L_k)$, besides the second-order ones $(\dot{L}_i-\dot{L}_{i'})$, $(\ddot{L}_i-\ddot{L}_{j})$, $(\dot{L}_i^2-\dot{L}_{j}^2)$.

\begin{table}[h]
	\centering
	\caption{The three relevant combinations of the TDI coefficients for the twelve-link geometric TDI combination.}
	\newcommand{\tabincell}[2]{\begin{tabular}{@{}#1@{}}#2\end{tabular}}
	\renewcommand\arraystretch{2}
	\begin{tabular}{|c|c|}
		\hline
		\hline
	TDI combination & $\tilde{P}_i+\tilde{P}_{(i+1)'}$\\
		\hline
		$[\alpha]_1^{12}$&$\left( 6,  0,   -6 \right)(iu)^2$ \\
		\hline
		$[\alpha]_2^{12}$&$\left( 2,   0,   -2 \right)(iu)^2$ \\
		\hline
		$[\alpha]_3^{12}$&$\left( 0,  0,   0 \right)(iu)^3$ \\
		\hline
		\hline
	\end{tabular}
	\label{12linkexpression}
\end{table}

\begin{table}
	\centering
	\caption{The four relevant combinations of the TDI coefficients for the fourteen-link TDI combinations.}
	\newcommand{\tabincell}[2]{\begin{tabular}{@{}#1@{}}#2\end{tabular}}
	\renewcommand\arraystretch{2}
	\begin{tabular}{|c|c|}
		\hline
		\hline
		TDI combination & $\tilde{P}_i+\tilde{P}_{(i+1)'}$ \\
		\hline
		$[U]_1^{14}$&$\left(4,     0,    -4\right)(iu)^2$\\
		 \hline
		$[U]_2^{14}$&$\left(4,     0,    -4\right)(iu)^2$\\
	    \hline
		$[EP]_1^{14}$&$\left(4,     0,    -4\right)(iu)^2$\\
	    \hline
		$[EP]_2^{14}$&$\left(0,     0,    0\right)(iu)^3$\\
		\hline
		\hline
	\end{tabular}
	\label{14linkexpression}
\end{table}

\begin{table}
	\centering
	\caption{The nine relevant combinations of the TDI coefficients for the sixteen-link modified second-generation TDI combinations.}
	\newcommand{\tabincell}[2]{\begin{tabular}{@{}#1@{}}#2\end{tabular}}
	\renewcommand\arraystretch{2}
	\begin{tabular}{|c|c|}
		\hline
		\hline
		TDI combination & $\tilde{P}_i+\tilde{P}_{(i+1)'}$ \\
		\hline
		$[X]_1^{16}$&$\left( 16,     0,   -16 \right)(iu)^2$ \\
		\hline
		$[X]_2^{16}$&$\left( 8,     0,    -8 \right)(iu)^2$ \\
		\hline
		$[U]_1^{16}$&$\left( 12,    -6,    -6 \right)(iu)^2$\\
		\hline
		$[U]_2^{16}$&$\left( 8,    -4,    -4 \right)(iu)^2$\\
		\hline
		$[U]_3^{16}$&$\left( 4,    -2,    -2 \right)(iu)^2$\\
		\hline
		$[E]_1^{16}$&$\left(  4,     0,    -4 \right)(iu)^2$\\
		\hline
		$[E]_2^{16}$&$\left( 2,     0,    -2 \right)(iu)^2$\\
		\hline
		$[P]_1^{16}$&$\left( 0,     4,    -4 \right)(iu)^2$\\
		\hline
		$[P]_2^{16}$&$\left(0,    2,    -2 \right)(iu)^2$\\
		\hline
		\hline
	\end{tabular}
	\label{16modlinkexpression}
\end{table}

\begin{table}
	\centering
	\caption{The thirteen relevant combinations of the TDI coefficients for the sixteen-link second-generation TDI combinations.
	These thirteen second-generation TDI combination solutions can be classified by their first-generation counterparts.
    Here, the solutions denoted by $*$ correspond to the combinations whose laser noise residuals possess the first-order contributions of the form $\dot{L}_i(L_j-L_k)$, besides the second-order ones $(\dot{L}_i-\dot{L}_{i'})$, $(\ddot{L}_i-\ddot{L}_{j})$, $(\dot{L}_i^2-\dot{L}_{j}^2)$.
    }
	\newcommand{\tabincell}[2]{\begin{tabular}{@{}#1@{}}#2\end{tabular}}
	\renewcommand\arraystretch{2}
	\begin{tabular}{|c|c|}
		\hline
		\hline
		TDI combination & $\tilde{P}_i+\tilde{P}_{(i+1)'}$ \\
		\hline
		$[U]_4^{16}$&$\left( 16,   -6,  -10 \right)(iu)^2$\\
		\hline
		$[U]_5^{16}$&$\left( 8,  -2,   -6 \right)(iu)^2$\\
		\hline
		$[U]_6^{16}$&$\left( 8,  -2,   -6 \right)(iu)^2$\\
		\hline
		$[PE]_1^{16}$&$\left( 2,   0,   -2 \right)(iu)^2$\\
		\hline
		$[PE]_2^{16}$&$\left( 2,   0,  -2 \right)(iu)^2$\\
		\hline
		$[PE]_3^{16}$&$\left(  6,   0,  -6 \right)(iu)^2$\\
		\hline
		$[PE]_4^{16}$&$\left( 0,   6,  -6 \right)(iu)^2$\\
		\hline
		$[PE]_5^{16}$&$\left(  0,   2,  -2 \right)(iu)^2$\\
		\hline
		$[PE]_6^{16}$&$\left( 0,   2,  -2 \right)(iu)^2$\\
		\hline
		$[PE]_7^{16}$&$\left( 0,   2,  -2 \right)(iu)^2$\\
		\hline
		$[PE]_8^{16}$&$\left( 0,   2,  -2 \right)(iu)^2$\\
		\hline
		$[PE]_9^{16}*$&$\left( 0,   2,  -2\right)(iu)^2$\\
		\hline
		$[PE]_{10}^{16}*$&$\left( 0,   2,  -2 \right)(iu)^2$\\
		\hline
		\hline
	\end{tabular}
	\label{16linkexpression}
\end{table}

\begin{table}
	\centering
	\caption{The sixteen relevant combinations of the TDI coefficients for the sixteen-link second-generation TDI combinations.
	These sixteen second-generation combinations are generic and do not pertain to any existing class.
 Here, the solutions denoted by $*$ correspond to the combinations whose laser noise residuals possess the first-order contributions of the form $\dot{L}_i(L_j-L_k)$, besides the second-order ones $(\dot{L}_i-\dot{L}_{i'})$, $(\ddot{L}_i-\ddot{L}_{j})$, $(\dot{L}_i^2-\dot{L}_{j}^2)$.}
	\newcommand{\tabincell}[2]{\begin{tabular}{@{}#1@{}}#2\end{tabular}}
	\renewcommand\arraystretch{2}
	\begin{tabular}{|c|c|}
		\hline
		\hline
		TDI combination & $\tilde{P}_i+\tilde{P}_{(i+1)'}$\\
		\hline
		$[T]_{1}^{16}$&$\left(  2,     0,    -2 \right)(iu)^2$\\
		\hline
		$[T]_{2}^{16}$&$\left( 2,     0,    -2 \right)(iu)^2$\\
		\hline
		$[T]_{3}^{16}$&$\left( 8,     0,    -8 \right)(iu)^2$\\
		\hline
		$[T]_{4}^{16}*$&$\left( 2,     0,    -2 \right)(iu)^2$\\
		\hline
		$[T]_{5}^{16}$&$\left( 6,     0,    -6 \right)(iu)^2$\\
		\hline
		$[T]_{6}^{16}$&$\left( 8,     0,    -8 \right)(iu)^2$\\
		\hline
		$[T]_{7}^{16}$&$\left( 10,     0,   -10 \right)(iu)^2$\\
		\hline
		$[T]_{8}^{16}$&$\left( 2,     0,    -2 \right)(iu)^2$\\
		\hline
		$[T]_{9}^{16}*$&$\left( 0,     0,     0 \right)(iu)^3$\\
		\hline
		$[T]_{11}^{16}$&$\left( -2,     2,     0 \right)(iu)^2$\\
		\hline
		$[T]_{12}^{16}$&$\left( 0,     0,     0 \right)(iu)^3$\\
		\hline
		$[T]_{14}^{16}*$&$\left( 0,     0,     0 \right)(iu)^3$\\
		\hline
		$[T]_{15}^{16}$&$\left( 0,     0,     0 \right)(iu)^3$\\
		\hline
		$[T]_{16}^{16}*$&$\left( 0,     2,    -2 \right)(iu)^2$\\
		\hline
		$[T]_{18}^{16}*$&$\left( 0,     0,     0\right)(iu)^3$\\
		\hline
		$[T]_{19}^{16}*$&$\left( 0,     0,    0 \right)(iu)^3$\\
		\hline
		\hline
	\end{tabular}
	\label{16linkexpression17}
\end{table}

\acknowledgments

This work is supported by the National Key R$\&$D Program of China under Grant No.2022YFC2204602, the Natural Science Foundation of China (Grant No. 12405060), the Postdoctoral Science Foundation of China (Grant No.2022M711259).

\bibliographystyle{unsrt}
\bibliography{reference_wang}{}

\end{document}